\documentclass[preprint,review,12pt,3p]{elsarticle}
\usepackage{pdfpages} 
\usepackage{pgffor} 

\makeatletter
\AtBeginDocument{\let\LS@rot\@undefined}
\makeatother

\def\supplementfilename{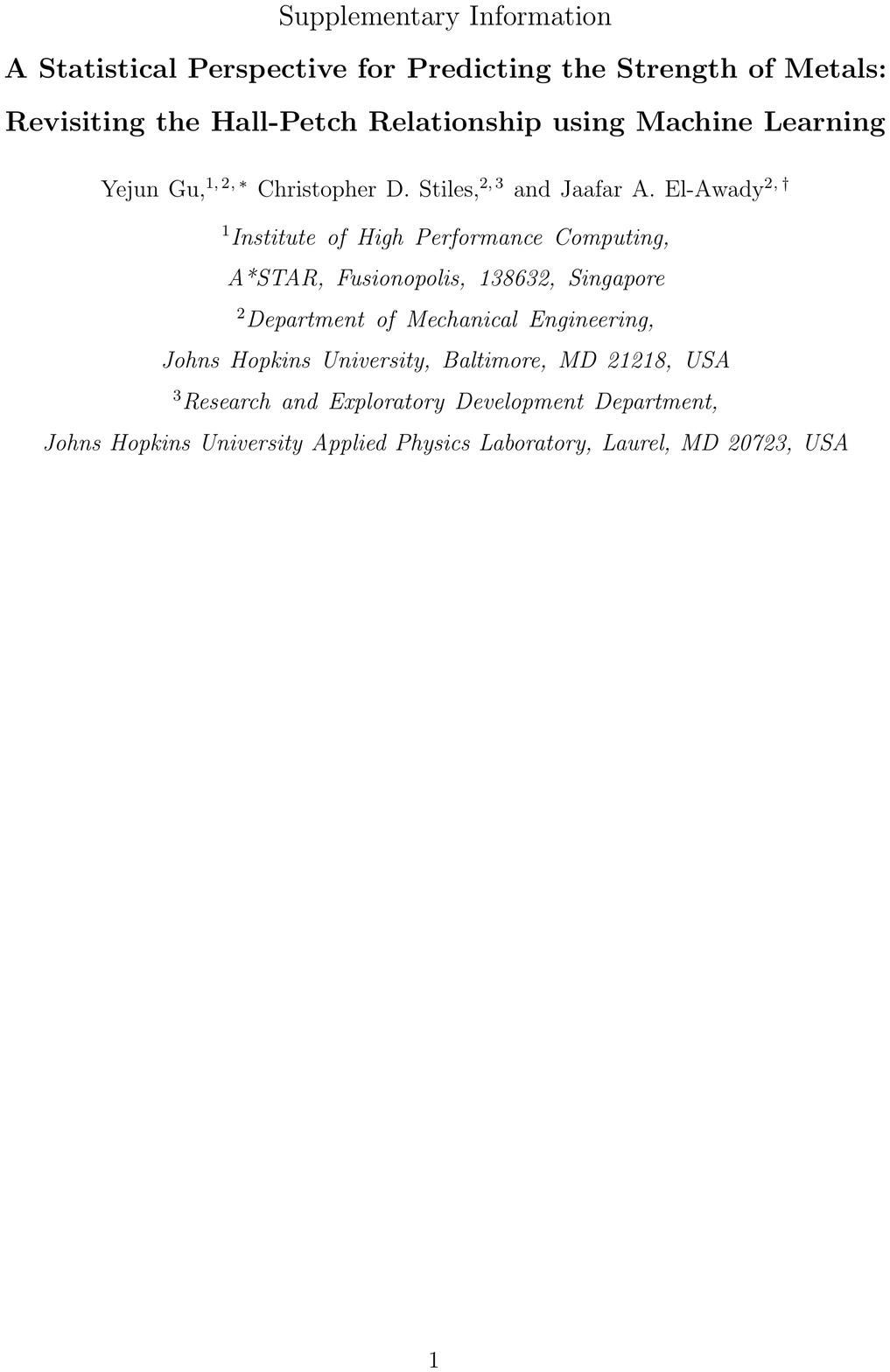}

\pdfximage{\supplementfilename}
\def\numbersupplementpages{\the\pdflastximagepages}

\usepackage{amssymb,amsmath}
\usepackage{siunitx}

\journal{}

\begin{document}

\begin{frontmatter}



\title{A Statistical Perspective for Predicting the Strength of Metals: Revisiting the Hall-Petch Relationship using Machine Learning}

\author[a,b]{Yejun Gu\corref{cor1}}
\ead{yejun_gu@ihpc.a-star.edu.sg}
\author[b,c]{Christopher D. Stiles} 
\author[b]{Jaafar A. El-Awady\corref{cor1}}
\ead{jelawady@jhu.edu}
\cortext[cor1]{Corresponding authors}

\affiliation[a]{organization={Institute of High Performance Computing, A*STAR},
                city={Singapore},
                postcode={138632}, 
                country={Singapore}}
\affiliation[b]{organization={Department of Mechanical Engineering, Johns Hopkins University},
                city={Baltimore},
                postcode={21218},
                state={MD},
                country={USA}}
\affiliation[c]{organization={Research and Exploratory Development Department, Johns Hopkins University Applied Physics Laboratory}, 
                cite={Laurel},
                postcode={20723},
                state={MD},
                country={USA}}

\begin{abstract}
The mechanical properties of a material are intimately related to its microstructure. This is particularly important for predicting mechanical behavior of polycrystalline metals, where microstructural variations dictate the expected material strength.  Until now, the lack of microstructural variability in available datasets precluded the development of robust physics-based theoretical models that account for randomness of microstructures. To address this, we have developed a probabilistic machine learning framework to predict the flow stress as a function of variations in the microstructural features. In this framework, we first generated an extensive database of flow stress for a set of over a million randomly sampled microstructural features, and then applied a combination of mixture models and neural networks on the generated database to quantify the flow stress distribution and the relative importance of microstructural features. The results show excellent agreement with experiments and demonstrate that across a wide range of grain size, the conventional Hall-Petch relationship is statistically valid for correlating the strength to the average grain size and its comparative importance versus other microstructural features. This work demonstrates the power of the machine-learning based probabilistic approach for predicting polycrystalline strength, directly accounting for microstructural variations, resulting in a tool to guide the design of polycrystalline metallic materials with superior strength, and a method for overcoming sparse data limitations.
\end{abstract}




\end{frontmatter}


\section{Introduction}
\label{sec:sample1}
Discovering and designing materials with extraordinary properties is the ultimate goal of materials science. For structural materials, high strength is one of the most important mechanical properties, which can be enhanced by optimizing the microstructures~\cite{Dieter1961,Timoshenko1983}. Among all strengthening mechanisms (grain refinement, strain hardening, solid solution strengthening, precipitate strengthening, and grain boundary hardening, etc.), grain refinement is arguably one of the simplest and most effective ways to increase the material strength. It is commonly known that the flow stress, $\sigma_f$, at a given macroscopic plastic strain, $\varepsilon$, increases when the average grain size, $d_{ave}$, decreases. This size effect is usually expressed by an empirical-based power law relationship in the form~\cite{Armstrong2014,Cordero2016}:
\begin{align}
    \sigma_f(\varepsilon)=\sigma_0(\varepsilon)+k_0(\varepsilon)\cdot d_{ave}^{-n},
    \label{eq:power_law}
\end{align}
where $\sigma_0$ and $k_0$ are fitting constants that depend on the chemistry, microstructure, and strain, while $n$ is the grain size exponent, and $\sigma_0$ is approximately the flow stress of the coarse-grained and untextured polycrystal. When $n=0.5$, Eq.~\eqref{eq:power_law} becomes the conventional Hall-Petch relationship, which was postulated by Hall and Petch in the early 1950's~\cite{hall1951,petch1953}. Experimental results of metallic materials generally support the empirical validity of Eq.~\eqref{eq:power_law}~\cite{Conrad1963, Hirth1972, Hansen2004, Cordero2016}. It should be noted that here $\varepsilon$ denotes the plastic strain, which is the subtraction of the total strain and the elastic strain. Under the tensile loading conditions, the elastic strain is $\sigma_f(\varepsilon)/E$, where $E$ is the Young's modulus. Thus, the total strain, $\varepsilon_{tot}$, is related to the plastic strain by $\varepsilon_{tot}=\varepsilon+\sigma_f(\varepsilon)/E$.

Many theoretical models have been proposed in literature to explain the physics that may be responsible for such a power law relationship between the average grain size and the polycrystal flow stress, including the pile-up model~\cite{hall1951,Cottrell1978}, the composite model~\cite{Thompson1973,Meyersm1982}, and the grain boundary source model~\cite{Li1963,Bata2004}. Numerical computational methods, such as molecular dynamics simulations~\cite{Jeon2011,Xu2018}, discrete dislocation dynamics (DDD) simulations~\cite{Biner2002,Biner2003,Balint2005,Balint2008,DeSansal2009,Levine2006,El-Awady2015}, and continuum crystal plasticity models~\cite{Acharya2000,Aifantis2006,Ohno2007,Haouala2018,Haouala2020} have also been used to quantify the underlying deformation mechanisms that may dominate such a power-law relationship. Nevertheless, it is noteworthy that the reported values in literature for the grain size exponent, $n$,  based on experiments and simulations are largely not consistent, even for the same material when studied by different groups. For example, various experimental fitting values for the grain size exponents were reported~\cite{baldwin1958,raj1986,Narutani1991,aghaie2012}. For Cu, Jiang et al. reported a value of 0.5 at low strains but not for high strains~\cite{jiang2013}, while a wider range, between 0.22 and 1.37, was found by Li et al.~\cite{Li2016}.  
Additionally, DDD simulations show that the grain size exponent varies from $0.4$ to $1$~\cite{von2001dislocation,Biner2002,Biner2003,Yellakara2014,Jiang2019}, which strongly depends on dislocation density and strain level.

The reported values for $\sigma_0$ also vary considerably in literature. For example, in $\alpha$-iron, the maximum reported value of $\sigma_0$ is up to two times as large as the minimum one~\cite{Conrad1960,Tjerkstra1961,Jago1986}. Moreover, the reported strain-dependency of $k_0$ on the strain level in literature lacks consistency. In some experiments, $k_0$ was reported to be constant (iron~\cite{Tjerkstra1961} and copper~\cite{Meyers1995}) or decrease with increasing strain (copper~\cite{Meyers1995} and steels~\cite{Tsuchida2008}). In other studies, $k_0$ was observed to first decrease then slightly increase with strain (nickel and steel~\cite{Feaugas2003} and iron~\cite{Jago1986}). 

This large discrepancy in the reported parameters was attributed to variations in microstructure (e.g., initial dislocation density, penetrability of grain boundaries (GBs), and grain size distributions), as well as the competition between different deformation mechanisms that control the flow stress. On the other hand, other studies have suggested that the Hall-Petch model is an artefact of faulty data analysis~\cite{Dunstan2013,Li2016} that the fitting parameters are sensitive to the error in determining the grain size. Their Bayesian meta-analysis suggested that the strength could be proportional to either $\ln{d}\cdot d^{-1}$ or $d^{-1}$, with $d$ being the average grain size. 

Due to the complications associated with characterizing the microstructural features and the wide spread in the reported flow stresses for different metals, a dataset that contains a large number of possible configurations is needed to sufficiently account for the effect of randomness in microstructures and gain a comprehensive understanding of the relationship between the flow stress and the microstructural features for any given material system. However there is scarcity in the available experimental and simulation datasets in the literature, with only a hundred data points at most for any given material. This lack of data represents a major barrier for drawing any conclusive statements about the proper fitting law for the flow stress or the dependence of the parameters on the strain magnitude or other microstructural features. {Recent advances in data driven approaches (e.g., statistical machine learning (ML), and data mining) have gained increasing interest to enable better strength predictions for metals and alloys. In the past decade, deterministic ML techniques (including support vector machines, surrogate models and artificial neural networks) have been applied to steels~\cite{shen2019physical}, copper alloys~\cite{wang2019property}, aluminium alloys~\cite{li2020accelerated}, titanium alloys~\cite{reddy2015artificial}, nickel-based superalloys~\cite{deng2022intelligent}, and several types of multi-principal element alloys~\cite{wen2019machine, chang2019prediction,raabe2023accelerating}. Data mining is another data driven method that collects data and extracts physical insights from literature~\cite{rajan2005materials, rajan2015materials, zou2021integrating, pei2023toward}. One of the major issues with such studies is the lack of representative datasets that ultimately diminish the robustness of knowledge gained through the use of these data driven approaches. On the other hand, almost all those studies focused on the composition and phase design of alloys without considering the effects of microstructural features, such as dislocations and grain boundaries, since the role of defects in determining the materials strength are much more difficult to be quantified in experiments and simulations, as compared to the composition-strength relationship.} In addition, the point estimate (i.e., using a single value to represent the parameters of an unknown population) for the parameters in Eq.~\eqref{eq:power_law} in literature cannot be utilized to evaluate the uncertainty in flow stress associated with different microstructures, which is vital for accurate predictions of the material strength~\cite{derlet2015universal}. {To address these issues, we present a ML-aided framework for the probabilistic prediction of the flow stress of a polycrystalline metal as a function of microstructural features, in contrast to the traditional approach in materials modeling, which focuses on the prediction of the averaged flow stress~\cite{vdg2020}}. 

We first formulate a physics informed theoretical framework to predict the flow stress as a function of microstructural features in a randomized polycrystal sample to generate representative datasets. This allows us to investigate the applicability of the Hall-Petch relationship and derive an appropriate form to describe the dependency of grain size on strengthening. We then apply a machine learning framework that combines mixture models and neural networks on the generated dataset to derive the material flow stress distribution as a function of microstructural features. Finally, we systematically analyze the relative importance of the different microstructural features that control the size effects of the flow stress.

\section{Results}
In dislocation-mediated crystal plasticity, the polycrystal stress as a function of plastic strain and different microstructural features is formulated here based on the generalized size-dependent Taylor-strengthening law~\cite{El-Awady2015,Gu2020} and the weakest link mechanism~\cite{El-Awady2009,Gu2020}. In this model, the critical resolved shear stress (CRSS) of a specific slip system, $\tau_{CRSS}$, is determined as the minimum stress required for slip activation over all dislocation sources:
\begin{align}
\tau_{CRSS}=&\min\{\mu b/\lambda+\alpha\mu b\sqrt{\rho}+\tau_0:\nonumber\\
&\hspace{5em}\text{for all possible $\lambda$}\},
\label{eq:CRSS}
\end{align}
where $\mu$ is the shear modulus, $b$ is the magnitude of the Burgers vector, $\lambda$ is the effective length of a dislocation source on the slip system, $\alpha$ is a geometrical constant (typically around $0.5$), $\rho$ is the dislocation density, and $\tau_0$ is the friction stress~\cite{Gu2020}. In Eq.~\eqref{eq:CRSS}, $\mu b/\lambda$ is the stress required to overcome the dislocation line tension, while $\alpha\mu b\sqrt{\rho}$ is the forest hardening term. {The detailed description of the variables in Eq.~\eqref{eq:CRSS} can be found in Sec. S1 of SI Appendix.} 

The strength of an individual grain can then be computed as the smallest value of the minimum CRSSs of all slip systems in the grain and the crystal theoretical shear strength for dislocation nucleation, $\tau_{theor}$, divided by the corresponding Schmid factor, $m$, such that:
\begin{align}
\sigma_{grain}(\boldsymbol{\theta})=&\min\big\{\frac{\min\{\tau_{CRSS},\tau_{theor}\}}{m(\boldsymbol{\theta})}:\text{for all slip systems}\big\},
\label{eq:grain_str}
\end{align}
where $m(\mathbf{\theta})$ is the Schmid factor as a function of the grain orientation, $\mathbf{\theta}$, which can be defined by the Euler angle triplets and the Bunge convention (i.e., $(\psi_1, \psi_2, \psi_3) \in O = \{[0, 2\pi], [0, \pi], [0, 2\pi]\}$~\cite{Bunge2013,Rowenhorst2015}). 

Additionally, the dislocation density in a grain deformed to a plastic strain, $\varepsilon$, can be expressed as~\cite{El-Awady2015}
\begin{align}
\rho = \tilde{\rho}_0 + \frac{K_s\varepsilon}{bd},
\label{eq:rho}
\end{align}
\noindent where $K_s$ is a geometrical constant, $\tilde{\rho}_0$ is the initial dislocation density in the grain, and $d$ is the effective grain size, which is defined as the cube root of the grain volume.

Finally, consider a cuboid polycrystal divided into an arbitrary number of  sections each having an arbitrary thickness along the loading direction. Each section is composed of various through-thickness grains having different grain sizes. The polycrystal strength, $\sigma_f$, can thus be defined to be equal to the strength of the weakest section along the loading direction {(i.e., iso-strain condition)}, and the strength of each section is controlled by the strongest grain in that section {(i.e., iso-stress condition)}~\cite{Eastman2016,Gu2020}, thus,
\begin{align}
    \sigma_f=&
    \min[\max\left(\sigma_{grain}(\boldsymbol{\theta})~\text{in the section}\right):\nonumber\\
    &\hspace{5em}\text{along the loading direction}].
    \label{eq:sig_f}
\end{align}

This formulation for the polycrystal strength represented by Eqs~(\ref{eq:CRSS}) and~(\ref{eq:sig_f}), are based on experiments~\cite{Eastman2016} and simulations~\cite{El-Awady2015} that have been critically examined in Ref.~\cite{Gu2020}. 

\begin{figure}[!tbhp]
\centering
\includegraphics[width=13cm]{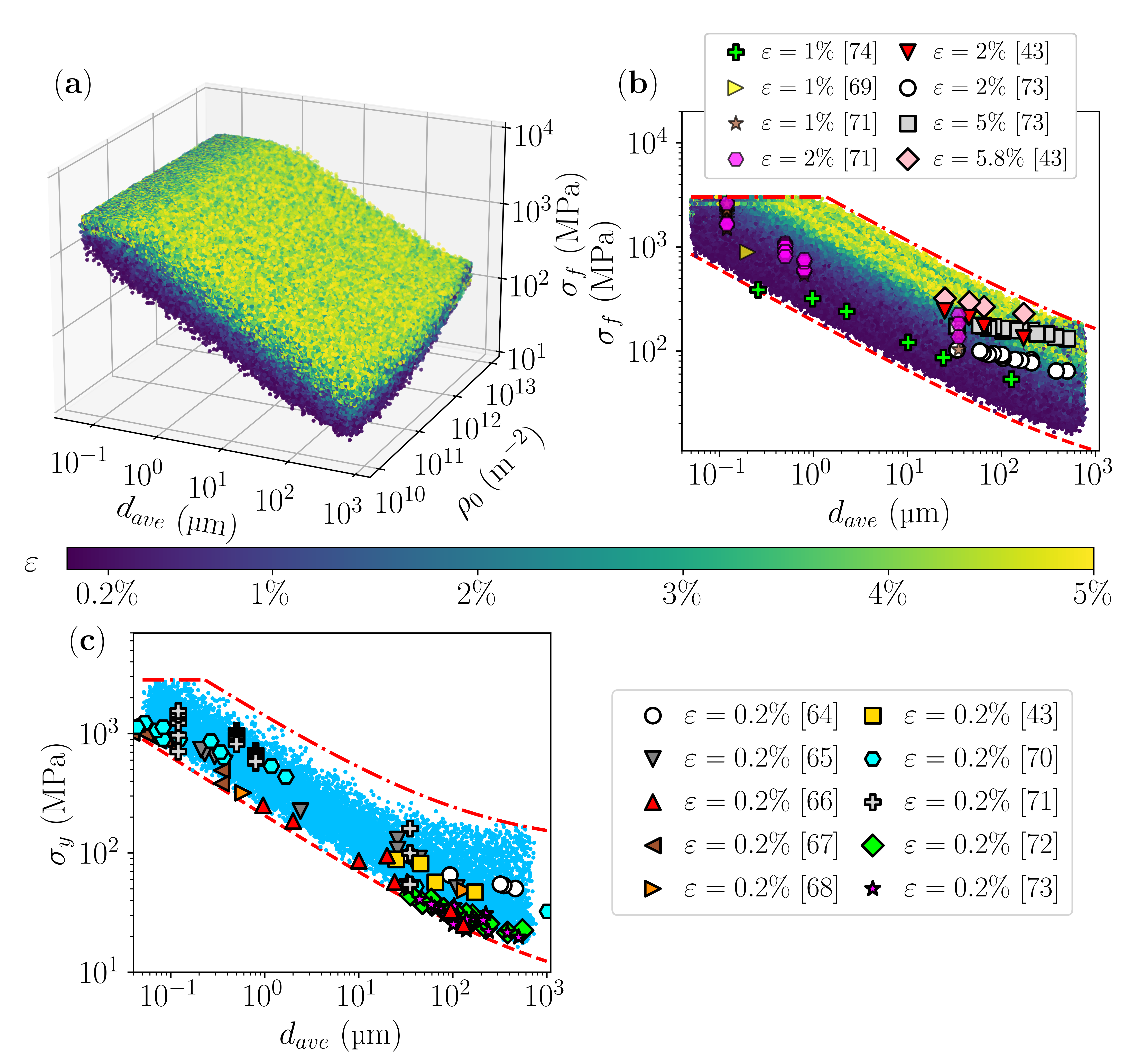}
\caption{An overview of theoretically predicted flow stresses calculated from Eqs.~(\ref{eq:grain_str}-\ref{eq:sig_f}). (a) 3D scatter plot showing the theoretically predicted flow stress, as a function of average grain size, dislocation density, and strain level. (b) The same data in (a) is collapsed on to a 2D scatter plot showing the theoretically predicted flow stress versus average grain size for different strain levels. (c) 2D scatter plot showing the theoretically predicted yield strength (flow stress at $0.2\%$ plastic strain), as a function of average grain sizes. The dashed line and the dash-dotted line in (b) are $\sigma_f^{(l)} =\SI{5}{\MPa}+\SI{190}{\MPa.\um ^{1/2}}/\sqrt{d_{ave}}$ and $\sigma_f^{(u)} =\max\{\SI{54}{\MPa}+\SI{3475}{\MPa.\um ^{1/2}}/\sqrt{d_{ave}},\SI{3018}{\MPa}\}$, respectively. While the dashed line and the dash-dotted line in (c) are $\sigma_y^{(l)} =\SI{6}{\MPa}+\SI{200}{\MPa.\um ^{1/2}}/\sqrt{d_{ave}}$ and $\sigma_y^{(u)} =\max\{\SI{113}{\MPa}+\SI{1305}{\MPa.\um ^{1/2}}/\sqrt{d_{ave}},\SI{2830}{\MPa}\}$, respectively. Experimental measurements for pure polycrystalline Ni from~\cite{Feaugas2003},~\cite{Floreen1969}-~\cite{Thompson1977} are shown in (b) and (c) for comparison.} 
\label{fig:overall}
\end{figure}

Here, a total of 1,199,946 face centered cubic (FCC) Ni polycrystalline cuboid samples with random grain sizes, grain numbers, grain orientations, and initial dislocation densities, were randomly generated and the flow stresses at different plastic strain levels (up to $5\%$, {where the dislocation-mediated plasticity is the dominant deformation mechanism}) for each case were calculated using the combination of Eq.~(\ref{eq:grain_str}-\ref{eq:sig_f}), which will be hereafter referred to as ``theoretically predicted data''. The details of the polycrystalline sample generation and the material properties are discussed in the Methods Section and Sec.~S1 of SI Appendix. The theoretically predicted data for all samples are summarized in Fig.~\ref{fig:overall}(a), and the results are shown to be in excellent agreement with experimentally reported results in the literature for pure Ni~\cite{Feaugas2003,Floreen1969, Wilcox1972,Thompson1975,Ebrahimi1999,Xiao2001,Schwaiger2003,Meyers2006,hollang2006,Keller2008,Keller2011, Thompson1977} (all data are from tensile tests under quasi-steady loading conditions, i.e., strain rate between \SI{1e-5}{s^{-1}} and \SI{1e-3}{s^{-1}}). The results show that when the average grain size increases, the theoretically predicted flow stresses decrease and are bounded by two curves that have a reciprocal square root dependency on the average grain size, namely, $\sigma_f^{(l)} =\SI{5}{\MPa}+\SI{190}{\MPa.\um ^{1/2}}/\sqrt{d_{ave}}$ and $\sigma_f^{(u)}=\max\{\SI{3018}{MPa},\SI{54}{\MPa}+\SI{3475}{\MPa.\um ^{1/2}}/\sqrt{d_{ave}}\}$. The upper bound at the theoretical strength $\sigma_{theor}=\SI{3018}{\MPa}$ when the average grain size, $d_{ave}$, is smaller than \SI{1.37}{\um}, represents the cases when the flow strength reaches the theoretical strength for dislocation nucleation. The value of $\sigma_{theor}$ is obtained from Eq.~\eqref{eq:sig_f} with the CRSS being $\tau_{theor}$, which is also equivalent to the theoretical shear strength of Ni, $\tau_{theor}=\SI{833}{\MPa}$, divided by the Taylor factor of the sample. Fig.~\ref{fig:overall}(a) and (b) also clearly demonstrate the effect of pre-straining, i.e., the flow stress generally increases with increasing strain, which is consistent with experimental observations~\cite{Feaugas2003, Keller2011} and supported by the strain hardening model~\cite{Narutani1991, El-Awady2015, Naik2020}. Furthermore, the theoretically predicted yield strength (i.e., the flow stress at 0.2\% offset) from all cases is shown in Fig.~\ref{fig:overall}(c). Similar to the flow stress data in Fig.~\ref{fig:overall}(b), all results are observed to be bounded by two curves inversely proportional to the square root of the average grain size: $\sigma_y^{(l)} =\SI{6}{\MPa}+\SI{200}{\MPa.\um ^{1/2}}/\sqrt{d_{ave}}$ and $\sigma_y^{(u)}=\max\{\SI{2830}{\MPa},\SI{113}{\MPa}+\SI{1305}{\MPa.\um ^{1/2}}/\sqrt{d_{ave}}\}$, respectively. 
 
Additionally, for coarse-grained samples with $d_{ave}>\SI{100}{\um}$, the upper limit of the strength data plateaus at $\approx$\SI{200}{\MPa} as shown in Fig.~\ref{fig:overall}(c). This upper bound is mainly controlled by the forest hardening term $\max\{\frac{\alpha\mu b\sqrt{\tilde{\rho}_0}+\tau_0}{\max\left[m(\boldsymbol{\theta}):~\text{for all slip systems}\right]}\}\approx\SI{190}{\MPa}$, as the dislocation slip in large grains are easily activated. This is also in agreement with 3D DDD simulation results~\cite{Yellakara2014}.

\begin{figure}[!tbhp]
\centering
\includegraphics[width=13cm]{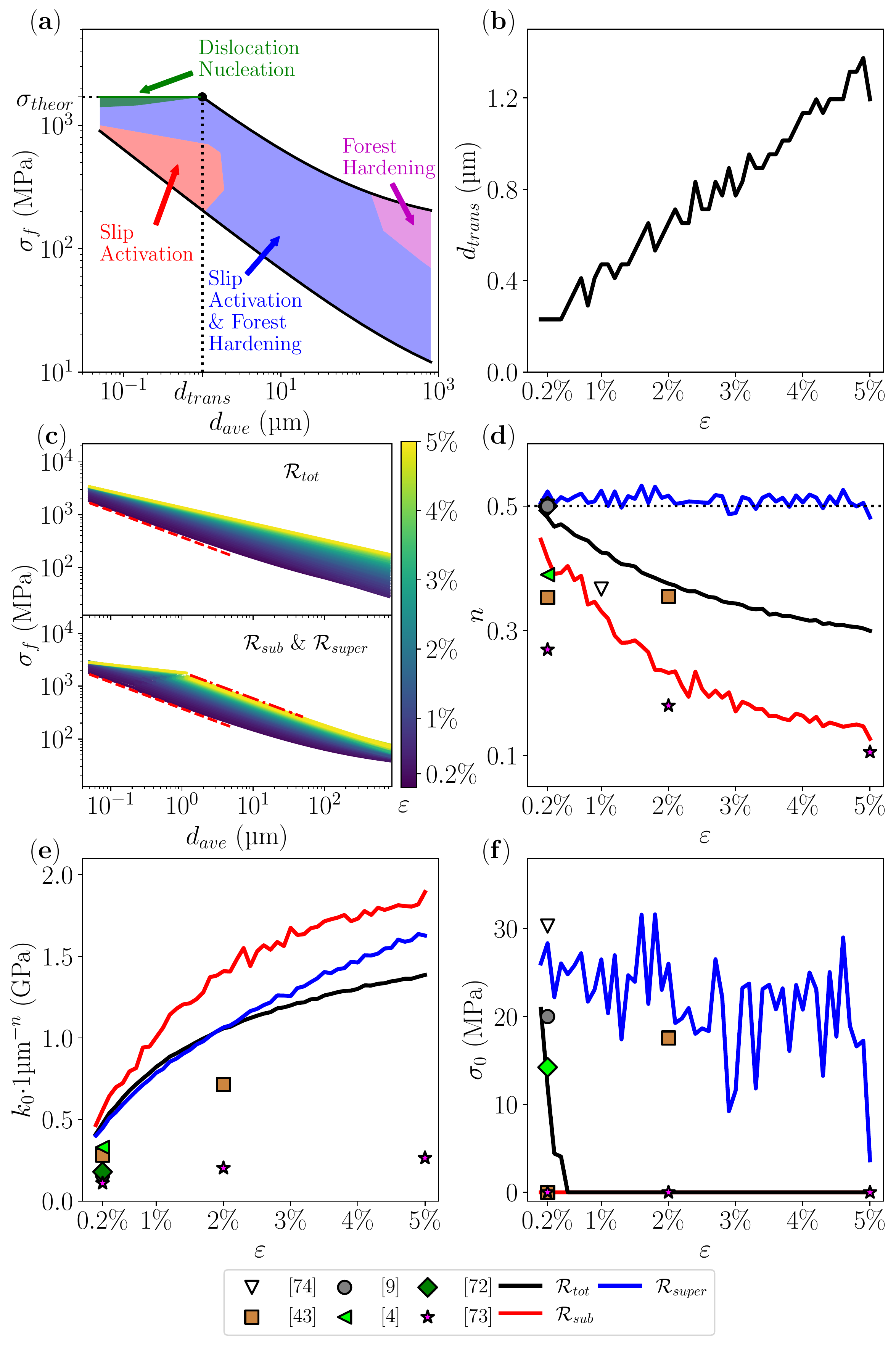}
\caption{The deformation mechanism map and the LSR analysis of deformation mechanisms dependencies, and their correspondence to the flow stresses. (a) A schematic overlaying the deformation mechanism map over the range of data shown in Fig.~\ref{fig:overall}. (b) The deformation transition grain size, $d_{trans}$, at different strain levels. (c) Constrained curve fitting of the flow stress for different ranges of average grain sizes, and the corresponding fitted values for (d) $n$, (e) $k_0\cdot \SI{1}{\um}^{-n}$, and (f) $\sigma_0$. The red dashed and dash-dotted lines having a slope of -0.5 in (c) are guide to the eye for the grain size exponent values. Fitting values reported in literature are also shown for comparison~\cite{Thompson1977, Feaugas2003, Hansen2004,  Keller2008, Keller2011, Cordero2016}.}
\label{fig:fit}
\end{figure}

As aforementioned, the different terms in Eq.~\eqref{eq:CRSS} represent the contributions from different deformation mechanisms to the flow stress. This is illustrated in Fig.~\ref{fig:overall} by the changes in the slope in different grain size regions. The domain of dominance for each deformation mechanism is shown schematically by the colored regions in Fig.~\ref{fig:fit}(a). It should be noted that these regions can overlap in response to different microstructural features even with the same average grain size. In the green region where the average grain size is small, the deformation is driven by dislocation nucleation from the grain boundaries or surfaces. Thus, the flow stress is equal to the theoretical stress for dislocation nucleation. On the other hand, when the average grain size, the strain, and the dislocation density are relatively low as shown by the red region, the flow stress is dominated by slip activation. However, when the average grain size, the strain, and dislocation density are high, shown by the magenta region, the flow stress is governed predominantly by dislocation forest hardening. Finally, in the intermediate region, the flow stress becomes a function of both slip activation and forest hardening. In Fig.~\ref{fig:fit}(a), a transition grain size, $d_{trans}$, can be defined as the point at which the governing deformation mechanism changes from dislocation nucleation to a combination of slip activation and forest hardening. The variation of $d_{trans}$ as a function of strain is shown in Fig.~\ref{fig:fit}(b). It is observed that $d_{trans}$ increases with increasing strain and reaches \SI{1.37}{\um} at $\varepsilon\approx5\%$. Since the transition grain size is interpreted as the turning point from a constant flow stress equal to $\sigma_{theor}$ (when $d<d_{trans}$) to a decreasing flow stress with increasing strain (when $d>d_{trans}$), at higher strains, the higher flow stress from strain hardening results in a larger transition grain size. However, when the grain size is above $d_{trans}$, the grains are large enough that slip activation and forest hardening overtake the dislocation nucleation to be the dominant mechanisms. Thus the flow stresses are below $\sigma_{theor}$ when $d_{ave}>d_{trans}$.

The best fit from the constrained least squares regression (LSR) analysis on Eq.~\eqref{eq:power_law} (subject to $\sigma_0>0$), linking the flow strength and average grain size, is shown in Fig.~\ref{fig:fit}(c). Fig.~\ref{fig:fit}(c) is a multiple line graph showing the fitting curves for the flow stress at different strain levels and in different ranges of grain size. The red dashed and dash-dotted lines having a slope of $-0.5$ are reference lines to compare the theoretically predicted data with the inverse square root law (i.e., the Hall-Petch relationship with a vanished $\sigma_0$). The upper panel in Fig.~\ref{fig:fit}(c) shows the fitting curves for the entire range of grain sizes under consideration here as denoted by $\mathcal{R}_{tot}$, while the lower panel shows the curves fitted in the sub-transition grain size range $d_{ave}<d_{trans}$ that is denoted by $\mathcal{R}_{sub}$ or the super-transition grain size range defined by $d_{ave}\geq d_{trans}$, denoted as $\mathcal{R}_{super}$, respectively. The fitting parameters of the curves shown in Fig.~\ref{fig:fit}(c) are presented in Fig.~\ref{fig:fit}(d-f) as a function of strain. Fig.~\ref{fig:fit}(d) shows that the grain size exponent in the super-transition grain size range ($\mathcal{R}_{super}$) varies between $0.482$ and $0.533$ for all strain levels. This demonstrates that the conventional Hall-Petch relationship is statistically valid for correlating the strength to the average grain size in this grain size range. On the other hand, the exponents in both $\mathcal{R}_{tot}$ and $\mathcal{R}_{sub}$ gradually decrease from $0.5$ towards $0.1$ with an increasing strain. This trend can be rationalized by the more significant contribution of the theoretical stress manifested by the increasing $d_{trans}$ with increasing strain at larger strain levels in $\mathcal{R}_{sub}$. For comparison, the exponents in the size effect power law as reported in literature, based on fitting to a significantly smaller number of data points, are typically between $0.2$ and $1$~\cite{Lefebvre2005,Dunstan2013,Dunstan2014, El-Awady2015}. The main reason for the discrepancy between the model predictions and that from the experimental predictions is the lack of statistically representative data from the experiments, which is overcome by the large set of data in the model.

Previous studies evaluating different forms of size effect laws from simulations or experiments (including both power law and non-power law) attempted to rationalize the wide range of exponents to the variations in deformation temperatures and the range of grain sizes investigated~\cite{Narutani1991,El-Awady2015}. Nevertheless, it is shown here that another major source for variations in the size effect laws is data scarcity in those previous studies, which failed to fully account for the likely variations in the statistics of the microstructural features (grain size distribution, grain orientation distribution, dislocation density, etc.) from one experiment to another. More specifically, the fitting may be sensitive to and compounded by the presence or absence of a few data points when the dataset size is not statistically satisfactory~\cite{Ryan2013}, such that there are many possible fitting curves to the data. On the other hand, with the unprecedentedly large size of representative data points examined here, the Hall-Petch relationship is shown to statistically describe the overall trend of the flow stresses as a function of average grain sizes that are larger than $d_{trans}$ (i.e., in $\mathcal{R}_{super}$). 

The fitting of the other two parameters, $k_0\cdot \SI{1}{\um}^{-n}$ (here $k_0$ is scaled by $\SI{1}{\um}^{n}$ to keep the unit consistency) and $\sigma_0$, from the current data points as a function of strain, are also shown in Fig.~\ref{fig:fit}(e) and (f), respectively. It is observed that $k_0\cdot \SI{1}{\um}^{-n}$ monotonically increases with increasing strain, emphasizing the pre-straining effect on the flow stress. Additionally,  despite some scatter in the variation of $\sigma_0$ with strain, $\sigma_0$ is always well below \SI{32}{\MPa} for the range of strains studied here. This indicates that the contribution of $\sigma_0$ in Eq.~\eqref{eq:power_law} to the flow stress can be considered negligible, as compared to the $k_0\cdot d_{ave}^{-n}$ term, except for large $d_{ave}$ and small $\varepsilon$. 


The overall trend of the flow stress as a function of the average grain size and strain level has been quantified above using the LSR method. Nevertheless, it is clear from Fig.~\ref{fig:overall} that the flow stress varies from its mean values by nearly an order of magnitude at the same average grain size and at the same strain level. {Hence, the deterministic prediction of the flow stress through Eq.~\eqref{eq:power_law} cannot describe the wide distribution in the flow stress associated with different microstructures, which should be accounted for in order to precisely predict the flow stress for materials discovery and optimization. We overcome this by considering the polycrystal flow stress as a random variable and obtaining its probability distribution, as an extension of the classical Hall-Petch relationship. Using the Gaussian mixture model, the probability density function (PDF) of the flow stress with the given average grain size and strain level can be described as a mixture of $k$ PDFs, where the $i$-th PDF follows a normal distribution with the mean $MEAN_i$ and the standard deviation $SD_i$,
\begin{align}
    p(\log_{10}\sigma_f|d_{ave},\varepsilon)=\sum_{i=1}^{k} \pi_i(d_{ave},\varepsilon)\mathcal{N}\big(MEAN_i(d_{ave},\varepsilon),SD_i^2(d_{ave},\varepsilon)\big).
    \label{eq:mixtures}
\end{align}
Here $\pi_i$ is the weight of the $i$-th PDF with $\sum_{i=1}^k \pi_i=1$. It should be noted that here the flow stress is scaled by the logarithmic transformation since the flow stress varies by multiple orders of magnitude in the range of average grain sizes chosen. Eq.~\eqref{eq:mixtures} can be extended to the case with further given microstructural features (e.g., average dislocation densities and volume weighted average Taylor factors). The parameters of the mixture components in Eq.~\eqref{eq:mixtures} are estimated using an effective probabilistic machine learning algorithm~\cite{bishop1994,Bishop2006}, termed the ``mixture density network'' (MDN). The trained MDN models are available on Zenodo, which repository address is provided in Sec.~\textit{Data and Code Availability}.}

\begin{figure}[!tbhp]
\centering
\includegraphics[width=13cm]{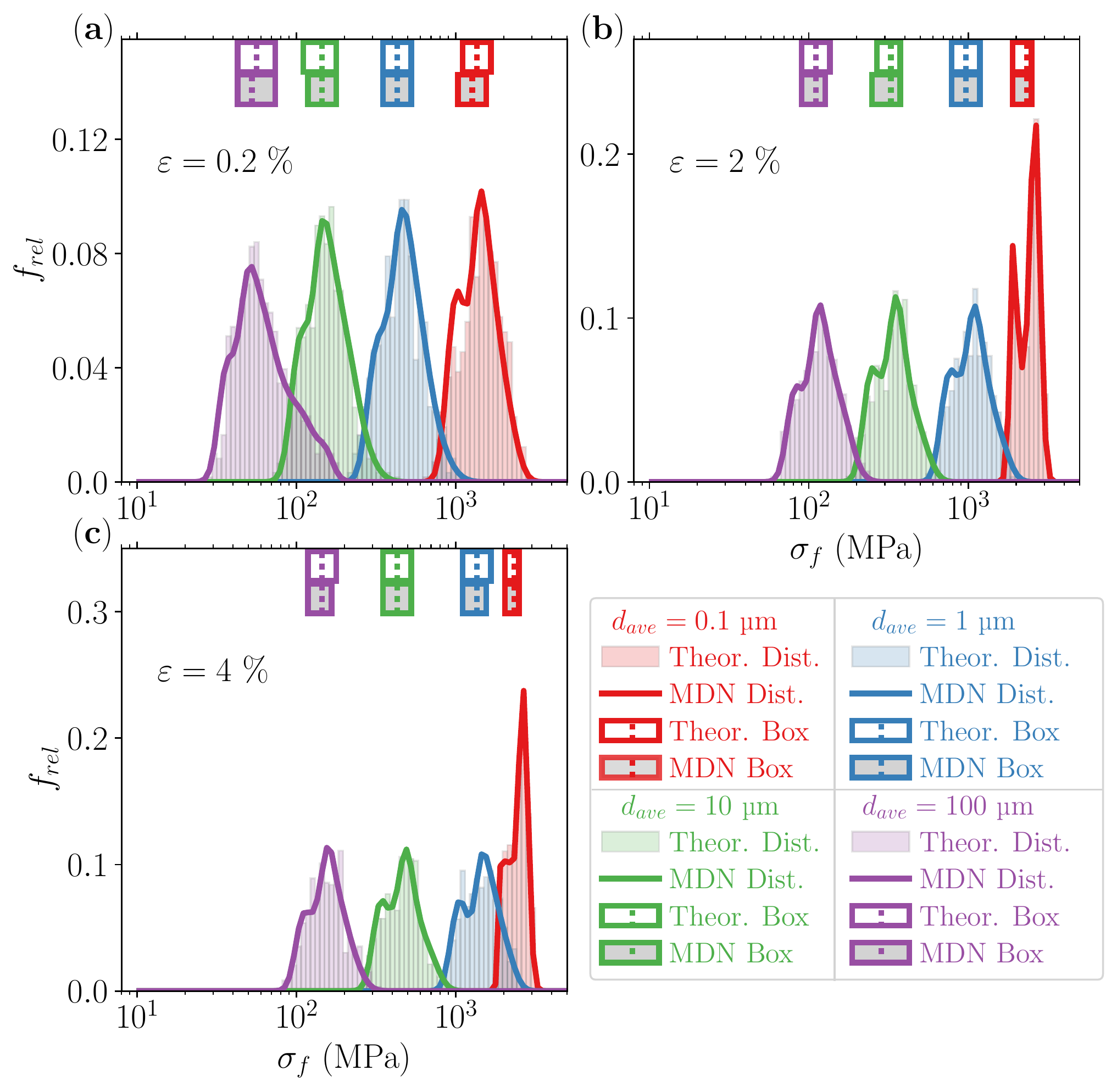}
\caption{The probability distributions of the theoretical and the MDN predicted flow stresses. The relative frequency, $f_{rel}$, and quartile box plot of the flow stresses with different average grain sizes for: (a) $\varepsilon=0.2\%$, (b) $\varepsilon=2\%$, and (c) $\varepsilon=4\%$. The box at the top of each subfigure extends from the Q1 (the leftmost vertical solid line of the box) to the Q3 (the rightmost vertical solid line of the box) values of the flow stresses while the vertical dotted line of the box represents the median. }
\label{fig:prob}
\end{figure}

The theoretical and MDN predicted flow stress distributions and quartile box plots for four different average grain sizes at three different strain levels are shown in Fig.~\ref{fig:prob}. The histograms and the lines are the distributions of the theoretical flow stresses and the MDN predicted flow stresses, respectively. Similarly, the white and grey boxes at the top of each subfigure represent the quartiles of the theoretically predicted and the MDN predicted flow stresses, respectively. The flow stresses at different grain sizes and strain levels are shown to follow a bimodal distribution. The existence of two local maxima of flow stress arises from the complicated relationship between the flow stress and microstructural features. The first quartile (Q1), second quartile (Q2) and third quartile (Q3) are defined as the cases under which 25\%, 50\% and 75\% of the data points are found when they are arranged in ascending order, respectively. While Fig.~\ref{fig:prob} only shows some representative results of the trained MDN model, all MDN predicted distributions are shown to be in excellent agreement with the theoretical predictions (see SI Appendix Sec.~S3 for all results). 

\begin{figure}[!tbhp]
\centering
\includegraphics[width=13cm]{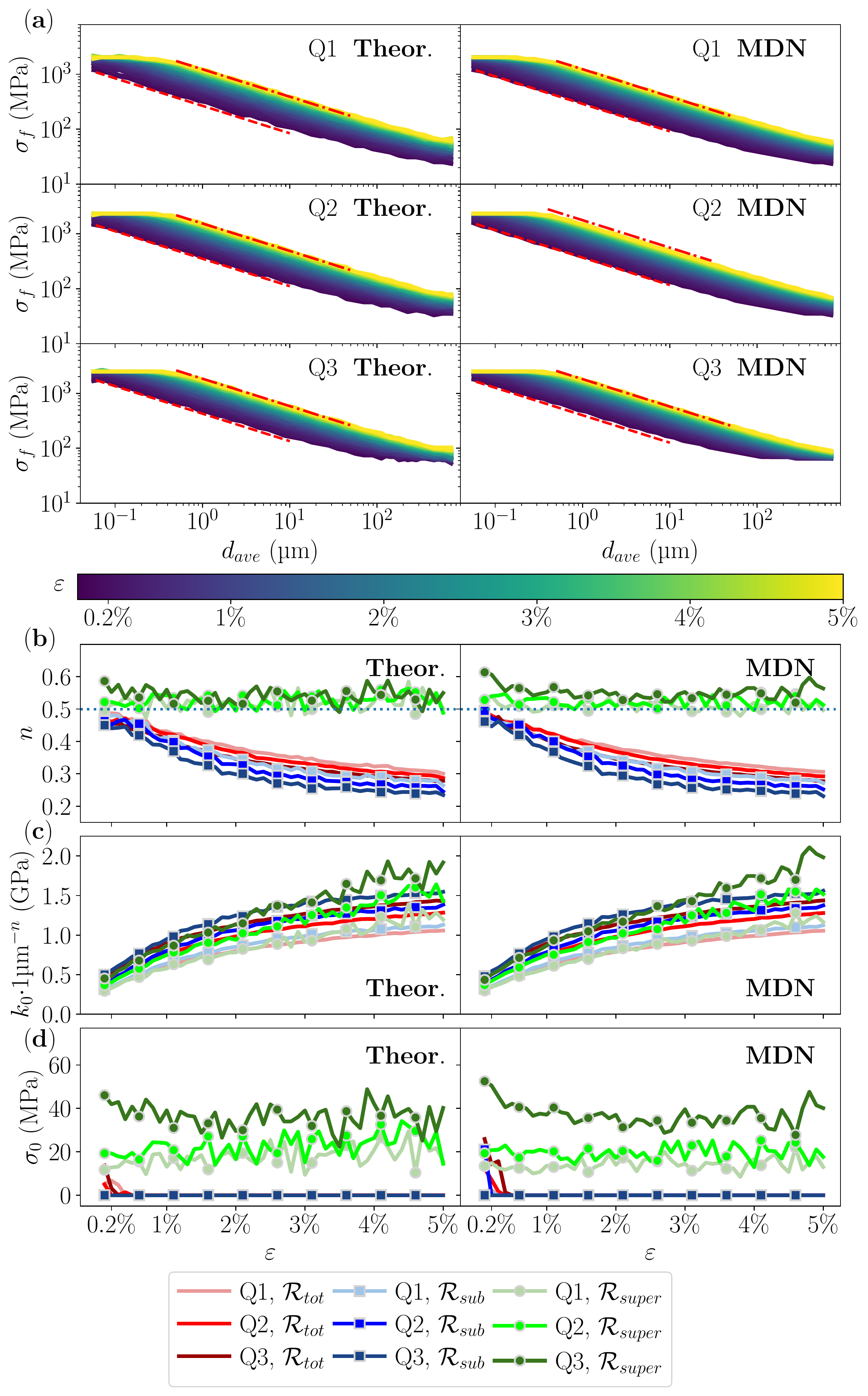}
\caption{Quartile analysis of the theoretical and MDN predicted flow stresses. (a) The quartiles of the theoretical (left) and the MDN (right) predicted flow stress for different average grain sizes at different strain levels, and the corresponding fitted values for (b) $n$, (c) $k_0\cdot \SI{1}{\um}^{-n}$ and (d) $\sigma_0$. The red dashed and dash-dotted lines having a slope of -0.5 in (a) are guide to the eye for the grain size exponent values.}
\label{fig:prob1}
\end{figure}

Further analysis of the quartiles of the theoretical and MDN predicted flow stresses is shown in Fig.~\ref{fig:prob1}. In Fig.~\ref{fig:prob1}(a), the lines are colored according to the strain levels. The left and right columns are from the theoretical and the MDN predicted distributions of the flow stresses, respectively. Similar to the LSR results, the quartile slopes from both the theoretical and the MDN predicted distributions are quite different for small ($\mathcal{R}_{sub}$) and large ($\mathcal{R}_{super}$) grain sizes, which is again due to the transition in the dominant deformation mechanisms. The curve fitting parameters, $n$, $k_0\cdot \SI{1}{\um}^{-n}$ and $\sigma_0$, of the quartiles, are shown in Fig.~\ref{fig:prob1}(b-d), respectively. From Fig.~\ref{fig:prob1}(b), the exponents in the coarse-grained materials (i.e., in $\mathcal{R}_{super}$) are around $0.5\sim0.6$, indicating a good alignment of the curves with $1/\sqrt{d_{ave}}$ dependence. While in fine-grained materials (i.e., in $\mathcal{R}_{sub}$), the grain size exponent gradually decreases from $0.5$ to $0.2$ as the grain size increases. The values of $k_0\cdot \SI{1}{\um}^{-n}$ and $\sigma_0$ for the quartiles have the same trends as those from the LSR fitting curves shown in Figs~\ref{fig:fit}(e,f), i.e., $k_0\cdot \SI{1}{\um}^{-n}$ monotonically increases as strain increases, and $\sigma_0$ are relatively small and negligible. The analysis based on the probability distribution of flow stress leads to concluding that for materials having average grain sizes larger than $d_{trans}$, the Hall-Petch relationship holds for the spread of the flow stress data, and the applicability of the Hall-Petch relationship is not a coincidence for the average flow stress. 

In the above analysis, an assumption was made based on empirical prior knowledge that the average grain size and strain level are two important factors in determining the flow stress. This assumption can be examined by analyzing the feature importance based on the trained MDN models. In particular, a series of commonly used microstructural features are analyzed in this work: 1. the average grain size (termed as ``GS''); 2. the grain size distribution (termed as ``GS\_DIS'') defined by the average grain size, the smallest grain size, the largest grain size, and the standard deviation in the grain size; 3. the average Taylor factor (termed as ``GO''); 4. the grain orientation distribution (termed as ``GO\_DIS'') defined by the average Taylor factor (of each grain), the smallest Taylor factor, the largest Taylor factor, and the standard deviation in the Taylor factor; 5. the volume weighted average Taylor factor (termed as ``wGO''), where the applied weight is defined as the grain volume divided by the average grain volume; 6. the volume weighted average grain orientation distribution (termed as ``wGO\_DIS'') defined by the volume weighted average Taylor factor (of each grain), the smallest volume weighted Taylor factor, the largest volume weighted Taylor factor and the (volume weighted) standard deviation of Taylor factor;  7. plastic strain level (termed as ``STRAIN''); 8. average initial dislocation density (termed as ``DEN''); and 9. normalized thickness (termed as “NT”) defined as the averaged number of grains along one dimension in a cross section. 

\begin{figure}[!tbhp]
\centering
\includegraphics[width=13cm]{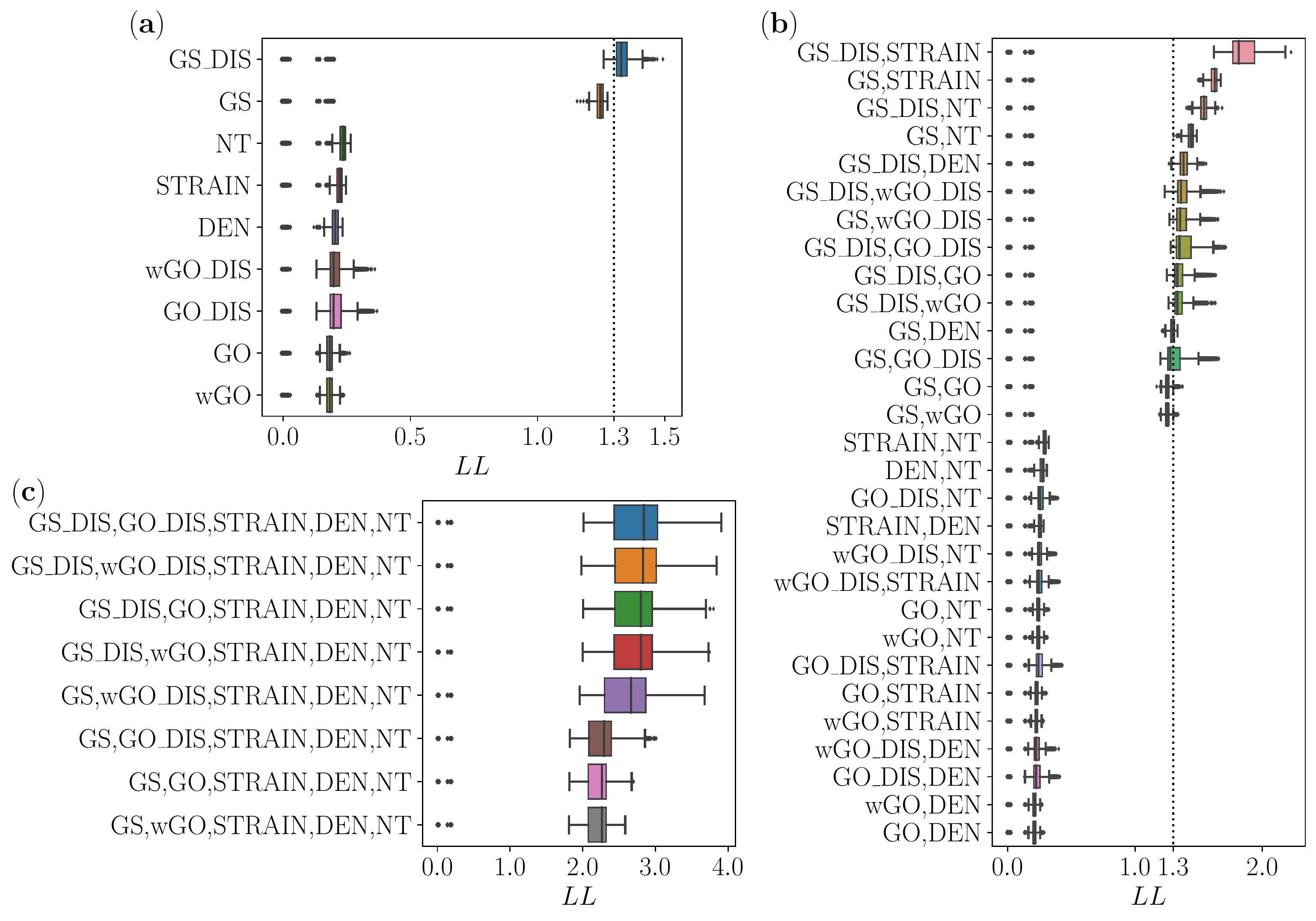}
\caption{Microstructural features for predicted flow stresses ranked by the log likelihood score for different feature combinations. (a) single type of features, (b) two types of features, and (c) all 5 types of features.}
\label{fig:feature}
\end{figure}

The contributions from those microstructural features on the flow stress are ranked by the median score of the mean log likelihood (denoted by ``LL'') as shown in Fig.~\ref{fig:feature}. It can be observed that the grain size related features, namely, the grain size distribution (``GS\_DIS'') and average grain size (``GS'') are the two most dominant factors in determining the flow stress. Since the feature average grain size (``GS'') is fully contained in the feature grain size distribution (``GS\_DIS''), the scores of the feature combinations with the grain size distribution are always higher than those of the same feature combination with the average grain size. According to the scores of a single type of features in Fig.~\ref{fig:feature}(a), each single feature not related to the grain size has average scores between $0.17$ and $0.23$. Even though the difference between the scores of different single features other than grain size is subtle, Fig.~\ref{fig:feature}(b) shows that scores associated with the dual features of the grain size/grain size distribution and strain are clearly higher than any other dual feature combination. Therefore, the features related to the grain size (``GS\_DIS'' or ``GS'') and strain are demonstrated to be the most important factors in controlling the flow stress. This analysis validates the widely adopted form of the Hall-Petch relationship as a function of average grain size and strain, as expressed by Eq.~\eqref{eq:power_law}. The grain size related features coupled with the normalized thickness also produce relatively high scores (ranking 3rd and 4th in Fig.~\ref{fig:feature}(b)), indicating the vital role of the normalized thickness in determining the flow strength~\cite{Eastman2016,Gu2020}. The scores from the grain size related features and the other two types of micrustructural features (average initial dislocation density, and grain orientation related features) are quite close, which sheds light on the complicated interdependence of deformation mechanisms related to dislocation density and grain orientation. When all five types of microstructural features are accounted for, the average scores are higher than 2, as demonstrated in Fig.~\ref{fig:feature}(c). This suggests that when the microstructural statistics are fully incorporated, the credibility of the MDN model prediction can be much improved. 

\section{Discussion and Conclusions}
In this work, the conventional Hall-Petch relationship alone is shown to be a statistically valid manifestation to describe the correlation between the flow stress and the average grain size that are larger than $d_{trans}$ (in $\mathcal{R}_{super}$), with the generation of over one million data points from Eq.~\eqref{eq:grain_str} and Eq.~\eqref{eq:sig_f}. Nevertheless, the large scatter in the observed stress values requires a quantification of the prediction uncertainty. To enhance the predictive accuracy, the Hall-Petch relationship must be complimented by a probabilistic description of flow stress distributions. This method characterizes the uncertainty and provides a method towards assessing the confidence level for a flow stress prediction. This approach provides the fundamental understanding and the statistics that are not always easily accessible through experiments alone. Furthermore, the trained MDN model can be used to generate results that are broader and more complete than the previous 70 years of experimental research without the use of any additional physical experiment. Therefore, this computational work has the potential to replace experimental testing by leveraging efficient while accurate strength prediction and deep insight on deformation mechanisms to meet material design targets on time and on cost. {Here we propose an approach that is centered on harnessing the power of data to accelerate materials design and its deployment. We combine the discovery of fundamental physical principles (i.e., a Newtonian paradigm) with machine learning approaches to enable scientific discoveries (i.e., a Keplerian paradigm). The approach enables producing accurate and representative datasets from physics-based models, and then effectively learn from the data to enable making accurate predictions. The problem we chose as a demonstrative case of this approach is the classical Hall-Petch relationship, which is arguably one of the most extensively used phenomenological relationships for predicting materials strength. However, this relationship has been debated for decades due to the scarcity of available experimental data. Our approach both overcomes this scarcity in data and harnesses the ability of machine learning to extract useful and novel physics.} Some important aspects in the current framework development will be discussed below.

\subsection*{Formulation of flow stress} In the formulation of the current theoretical model for the prediction of the FCC polycrystal flow stress, a few simplifications were employed, e.g., the effect of strain rate on the flow stress~\cite{messerschmidt2001,fan2021} was ignored. It has been shown not to significantly change the stress-strain curves of all nanocrystalline, ultra-fine crystalline and microcrystalline Ni when the strain rates are between \SI{3e-4}{s^{-1}} and \SI{3e-1}{s^{-1}}~\cite{Schwaiger2003}. Thus the results of this work can be directly applied for such a strain rate range. While the strain rate beyond this range possibly leads to a very different dislocation density evolution, which requires more systematic studies to improve the capability of this model for wider strain rate ranges. Furthermore, to apply this model from FCC materials to other crystal structures, some extra factors should be addressed, e.g., more complicated slip systems and anisotropy in slip resistance for different dislocation characters and on different slip systems in body-centred cubic and hexagonal close-packed materials. {In addition, it should be noted that this theoretical model was developed for homogeneous or weak texture materials at small strains, and some homogenization methods were used for microscopic plastic strains and non-equivalent grain growth/recovery. This approximation is accurate for predicting the homogenized properties at the macroscopic level (average behaviors of many grains, e.g., yield)~\cite{segurado2018,thool2020}. Nevertheless, the model can be improved to account for heterogeneous deformation, large strains and rotations, contact, and complex local strain histories to predict the macroscopic properties dominated by local microstructures (fatigue and creep crack initiation, crack propagation, etc.), which is beyond the current study.} Finally, the formulation developed here only applies to pure metals. To extend this formulation to predict the material flow stresses in alloys, additional terms associated with the strengthening mechanisms due to solid solutions or precipitates should be incorporated~\cite{Timoshenko1983}. 

\subsection*{Sources of uncertainty in the flow stress} There are two sources of the flow stress uncertainty in this work~\cite{chernatynskiy2013,vdg2020}: One is the deficiency due to an imperfection in calculations or a lack of knowledge of the physics law, which is referred to as the the epistemic (or systematic) uncertainty. The other is associated with the random microstructures of materials, which is referred to as the aleatoric (or statistical) uncertainty. The epistemic uncertainty can usually be eliminated to a sufficient degree by calibrating and validating the physics model using reliable results from experiments. In this work, this procedure has been performed during formulating the flow stress. While the aleatoric uncertainty is more significant with the information loss or the dimension reduction when the model is upscaled to a coarse-grained model, leading to the inherent randomness of the microstructures being not completely represented. The effect of the incomplete information of the microstructural randomness can be quantified using probability distributions, as presented in this work. Thus, the collection of more representative data would help characterize a more accurate probability distribution of flow stress, albeit this aleatoric uncertainty is irreducible and cannot be effectively mitigated. On the other hand, coarse-grained models are desired in order to overcome the inefficiency issue of atomistic resolution simulations to reach higher spatial and temporal scales, by decreasing the degrees of freedom. The capability and accuracy of coarse-grained models can be enhanced by properly choosing a series of homogenization methods or quantities that store and preserve key information to a large extent. For example, we have shown that the grain size distributions described by the average, standard deviation, lower bound and upper bound of grain sizes (i.e., GS\_DIS) can better predict the flow stress distribution than only taking average grain sizes, which highlights the paramount importance of appropriate dimension reduction in minimizing the aleatoric uncertainty. 

\subsection*{Treatment of data bias} The reliability of the knowledge extracted from data-driven approaches is highly dependent on the collection of a large assembly of unbiased data. The most practical way to reduce bias is to increase the dataset size. In this work, the one million simulation samples are equally generated from different microstructure distributions with different random seeds, such that there is no specific distribution being predominantly used and no skewed sample generated.  

\appendix
\section{Materials and Methods}
\subsection*{Formulation of polycrystal strength} Consider a cuboidal polycrystal under a loading parallel to the z-axis, as shown in Fig.~S1 of the supporting materials, with length $L$ (in the z-direction), thickness $T$ (in the y-direction), and width $D$ (in the y-direction). The polycrystal can be divided into multiple sections along the z-axis. Each section is composed of various grains. {Since GB sliding and grain rotations are only relevant for the prediction of strength in nanocrystalline materials (i.e., grain sizes $\leq$\SI{30}{\nm})~\cite{van2001,wei2004, Han2018}, these mechanisms are not accounted for in our model which focuses on grain sizes $\geq$\SI{50}{\nm}. To be consistent with this assumption, and for the sake of simplicity, the grains can be visualized as right prisms with all grain boundaries either being parallel or perpendicular to the loading direction. It should also be noted that the details of the grain shape (irregular grain boundaries and random grain boundary directions) can influence local strains and local dislocation density. However, these influences have been shown to average out when predicting the overall strength and when the strain level is below 8\%~\cite{feltham1957,sun2012, diehl2017, sun2019, ahn2021}.} The average grain size is defined as the cube root of the average grain volume, i.e., $d=\sqrt[3]{DTL/n_{grain}}$, where $n_{grain}$ is the total number of grains in the polycrystal. 

The following three steps are taken to mathematically formulate the strength in polycrystals~\cite{Gu2020}. 

\textbf{1. (Strength of a single grain)} For a slip system in a grain, the CRSS determined by the dislocation source that has the largest effective length $\lambda$, $\tau_{CRSS}=\{\mu b/\lambda+\alpha\mu b\sqrt{\rho}+\tau_0\}$, is capped at the theoretical strength related to thermal activation events of dislocations, $\tau_{theor}$. Thus, the grain strength is related to the CRSS, i.e., $\min\{\tau_{CRSS},\tau_{theor}\}$, by the Schmid factor, which is defined in Eq.~\eqref{eq:grain_str}. 

\textbf{2. (Strength of the section)} The strength of a given section is equal to that of the strongest grain in that section~\cite{Eastman2016}. 
\begin{align}
    \sigma_{sec}=\max\{\sigma_{grain}: \text{in section}\}.
    \label{eq:sec_str}
\end{align}

\textbf{3. (Strength of the sample)} The flow stress for the entire sample is determined by the weakest section:
\begin{align}
    \sigma_f=\min\{\sigma_{sec}: \text{along the loading direction}\}.
    \label{eq:sample_str}
\end{align}

The polycrystal strength is equal to the weakest section strength along the loading direction, while the section strength is defined as the strongest grain strength~\cite{Eastman2016,Gu2020}. Thus the polycrystal strength in Eq.~\eqref{eq:sig_f} becomes Eq.~\eqref{eq:sig_f0}.
\begin{figure*}[hbt!]
\begin{align}
    \sigma_f=\min\Bigg\{\max\Big[\min&\left(\frac{\min(\mu b/\lambda+\alpha\mu b\sqrt{\rho}+\tau_0,\tau_{theor})}{m(\boldsymbol{\theta})}: \text{for all slip systems}\right):\\\nonumber
    &\text{in section}\Big]: \text{along the loading direction}\Bigg\}.
    \label{eq:sig_f0}
\end{align}
\end{figure*}

Moreover, the dislocation storage rate in a grain with the size, $d$, under the quasi-static loading condition, is of the form~\cite{Conrad1967, Embury1971, Conrad2004, DeSansal2009, sun2019}: 
\begin{align}
    \frac{\partial \rho}{\partial \varepsilon}=\frac{K_s}{bd},
\end{align}
where $K_s$ is a constant for dislocation density evolution, which leads to a linear relationship (i.e., Eq.~\eqref{eq:rho}) between the dislocation density, $\rho$, and the macroscopic plastic strain, $\varepsilon$. In Eq.~\eqref{eq:rho}, $\tilde{\rho_0}$ for each grain is expected to be inversely proportional with the grain size~\cite{Conrad1967, Conrad2004}:
\begin{align}
\tilde{\rho}_0=\gamma\rho_{0}(1+D_{c}/d),
\label{eq:pc_rho1}
\end{align}
\noindent where $\gamma$ is a random variable ranging between 0.5 and 2 to account for the heterogeneity of the dislocation density in different grains, $D_{c}$ is a characteristic grain size, and $\rho_{0}$ is the mean initial dislocation density in the polycrystal.

It should be noted that the dislocation length distributions and dislocation density evolutions are actually dependent on the structures and properties of grain boundaries. Here for simplicity, also due to the lack of systematic studies on such grain boundary dependencies,  the dislocation length distributions and dislocation density evolutions are assumed to be the same in all grains. The implementation with a more complete description of the dislocation length distributions and dislocation density evolution at larger deformation stages (e.g., Kocks-Mecking model~\cite{mecking1981, Kocks2003}), will improve the accuracy of the flow stress prediction and will be presented elsewhere.

\subsection*{Mixture density networks}
\label{sec:mdn}
{A typical mixture density network is built from two basic components~\cite{bishop1994,Bishop2006}: a Gaussian mixture model and a neural network. The MDN possesses the benefits of both constituting models. The Gaussian mixture model can be used to estimate the distributions without imposing an ad hoc inductive biased assumption on the forms and functions of distributions. On the other hand, the neural network can calculate the parameters in the Gaussian mixture model by optimizing a given loss function. It confers excellent predictive capabilities on non-linear and complex relationships, as well as better generalization performance and better robustness~\cite{bishop1994,Bishop2006, schmidhuber2015}. MDN is chosen over other ML models as it is a well-established technique with significant flexibility to learn complex dynamics, and can be coupled with the Gaussian mixture method~\cite{graves2013,capes2017siri}. Many other ML model architectures, and configurations might also have the potential for such predictions and couplings. However, they are beyond the scope of this work.} The MDN models are implemented in Python 3 using Keras and TensorFlow libraries~\cite{Martin2019}, see more details in SI Appendix Sec.~S4. 

\section*{Data and Code Availability}

The dataset generated, the trained MDN models, and the codes used in this paper are available on Zenodo with the repository address: https://www.doi.org/10.5281/zenodo.7762663 
\section*{Acknoledgement}

{YG and JAE acknowledge financial support from the Office of Naval Research under grant \#N00014-18-1-2858. YG acknowledges financial support by the Structural Metal Alloys Program (A18B1b0061) of A*STAR in Singapore. JAE acknowledges support from U.S. National Science Foundation (NSF) award number DMR-1807708 and the Army Research Laboratory under Cooperative Agreement Number W911NF-22-2-0014. The authors also gratefully acknowledge internal financial support from the Johns Hopkins University Applied Physics Laboratory’s Internal Research \& Development (IR\&D) Program for funding portions of this work. The simulations were conducted at (1) the Advanced Research Computing at Hopkins (ARCH) core facility (rockfish.jhu.edu), which is supported by an NSF grant number OAC-1920103, (2) the Extreme Science and Engineering Discovery Environment (XSEDE) Expanse supercomputer at the San Diego Supercomputer Center (SDSC) through allocation TG-MAT210003, which is supported by National Science Foundation grant number ACI-1548562, and (3) the National Supercomputing Centre of Singapore through Allocation 13002696. We also thank Mr. Dylan Madisetti for helping publishing codes online. The views and conclusions contained in this document are those of the authors and should not be interpreted as representing the official policies, either expressed or implied, of the Army Research Office or the U.S. Government. The U.S. Government is authorized to reproduce and distribute reprints for Government purposes notwithstanding any copyright notation herein.}



    

\section*{Supplementary material}

Supplementary Information

# A Statistical Perspective for Predicting the Strength of Metals: Revisiting the Hall-Petch Relationship using Machine Learning

Yejun Gu,[1,2,*] Christopher D. Stiles,[2,3] and Jaafar A. El-Awady[2,†]

[1]*Institute of High Performance Computing,*
*A\*STAR, Fusionopolis, 138632, Singapore*

[2]*Department of Mechanical Engineering,*
*Johns Hopkins University, Baltimore, MD 21218, USA*

[3]*Research and Exploratory Development Department,*
*Johns Hopkins University Applied Physics Laboratory, Laurel, MD 20723, USA*

## S1. DATA GENERATION

The geometrical configuration of a cuboid polycrystal that is used for the theoretical calculation of the flow stresses is shown in Fig. S1. The sample is under a loading parallel to the z-axis, with length $L$ (in the z-direction), thickness $T$ (in the y-direction), and width $D$ (in the y-direction). The polycrystal can be divided into multiple sections along the z-axis. Each section is composed of various grains, as shown in Fig. S1. The grains are assumed to be right prisms and all grain boundaries (GBs) are either parallel or perpendicular to the loading direction. Thus, in the given cuboid polycrystalline sample, the grain structures can be generated using the two-dimensional Voronoi tessellation method on the xy-plane for quantitative evaluations of grain sizes [S1]. The effective grain size is determined as the cube root of the grain volume (i.e., the volume of the Voronoi cell) [S2]. It should be noted that the actual grain shapes are not accounted for in this model, since the grain shape effects are averaged out when considering microscopic properties such as yield or strength as a function of strain at early deformation stages [S3–S8]. Additionally, in each polycrystal, the grain size distribution is chosen to follow a truncated log-normal distribution, which is in strong agreement with experimental characterizations of polycrystalline materials [S3, S9–S11]. Different samples have varying scales and parameters of the grain size distribution, depending on the processing condition. Additionally, the grain orientations are represented by uniformly and randomly generated Euler angles. The weakest dislocation length in a grain is assumed to be proportional to $bd\sqrt{\rho}/\beta$ [S12], which is further assumed here to follow a uniform distribution between $0.5bd\sqrt{\rho}/\beta$ to $2bd\sqrt{\rho}/\beta$, with $\beta$ being a dimensionless constant, $b$ is the magnitude of the Burgers vector, $d$ is the grain size, and $\rho$ is the dislocation density in the grain as determined by Eq. (9) in the main text. All common material and the geometric properties for pure Ni used in all calculations are summarized in Table S1.

* Corresponding author

    yejun_gu@ihpc.a-star.edu.sg

† Corresponding author

    jelawady@jhu.edu

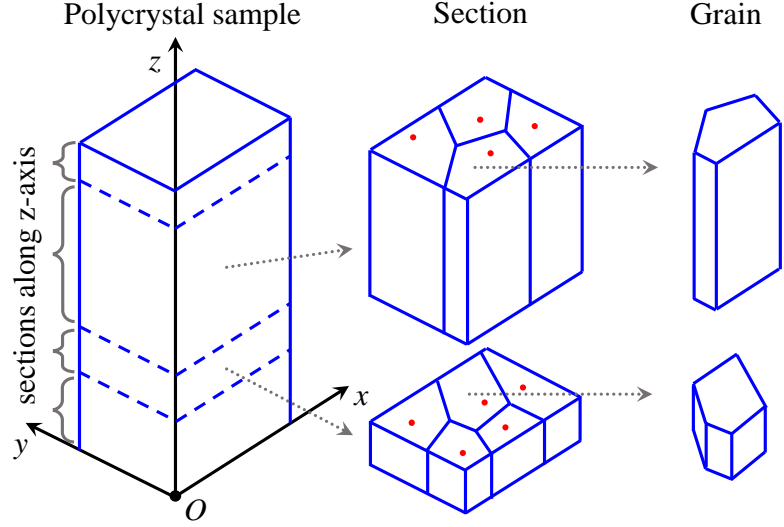

FIG. S1: Schematic representation of the polycrystal samples considered in this study. The grain structures are determined by the Voronoi tesselation method [S1], where the red dots represent the centroids of the Voronoi cells.

Also, the probability distributions of the microstructural features used in this work are summarized in Table S2.

TABLE S1: The material and the geometric properties for pure Ni used in all simulations

| Property | Symbol | Value |
|---|---|---|
| Shear modulus | $\mu$ | 76 GPa [S13] |
| Burgers vector magnitude | $b$ | 0.249 nm [S13] |
| Geometrical coefficient | $\alpha$ | 0.57 [S12] |
|  | $\beta$ | $1.76 \times 10^{-3}$ [S12] |
| Friction stress | $\tau_0$ | 2.5 MPa [S14] |
| Theoretical strength | $\tau_{theor}$ | 833 MPa [S13] |
| Number of slip systems | $N$ | 12 |
| Constant for dislocation density evolution | $K_s$ | 4.6 [S15] |
| Characteristic grain size | $D_c$ | 10 µm [S12] |

3

TABLE S2: The microstructural feature distribution used in all simulations

| Random variable | Symbol | Distribution |
| --- | --- | --- |
| Effective dislocation length | $\lambda$ | Uniform distribution |
| Euler angle triplets | $\tilde{\psi}_1, \tilde{\psi}_2, \tilde{\psi}_3$ | Uniform distributions |
| Grain size | $d$ | Truncated log-normal distribution |
| Number of grains in each direction | - | uniform distribution |

†: The Euler angle triples in this work follow the MTEX convection [S16], in order to invoke functions in MTEX. The MTEX orientations are defined as coordinate transformations from the crystal reference frame into the specimen reference frame, which is exactly the inverse of a Bunge orientation.

## S2. DISCUSSION ON FIGURE 1

There are four rogue experimental data points reported in literature (i.e., Refs.[66, 68, 74]) beyond the flow stress ranges from the model predictions. These data points are not plotted in Fig. 1(b) & (c), to avoid any confusion. The reasoning for this is that these data points are abnormal since they are associated with special microstructural configurations that are different from other published experimental data and the assumptions of our model predictions. The concerns with these data points and how the model may be tweaked to predict them are discussed below.

One experimental yield strength data point from Ref. [68], is below the predicted lower bound of the yield strength as predicted from our model and all other experimental measurements. This is due to the fact that this particular data point was from a sample that had a multimodal grain size distribution [68]. However, in the analysis of the main text a log-normal grain size distribution was chosen, which is believed to be representative for the majority of the other experimental data points in literature and shown in Fig. 1(c). Nevertheless, the current model predictions would in fact better match this abnormal data point if a bimodal grain size distribution is chosen, as shown in Fig. S2. In Fig. S2, a bimodal grain size distribution with the peaks at 0.2 µm and 12 µm is used, while other microstructural parameters are unchanged and microstructural features still follow the same probability distributions.

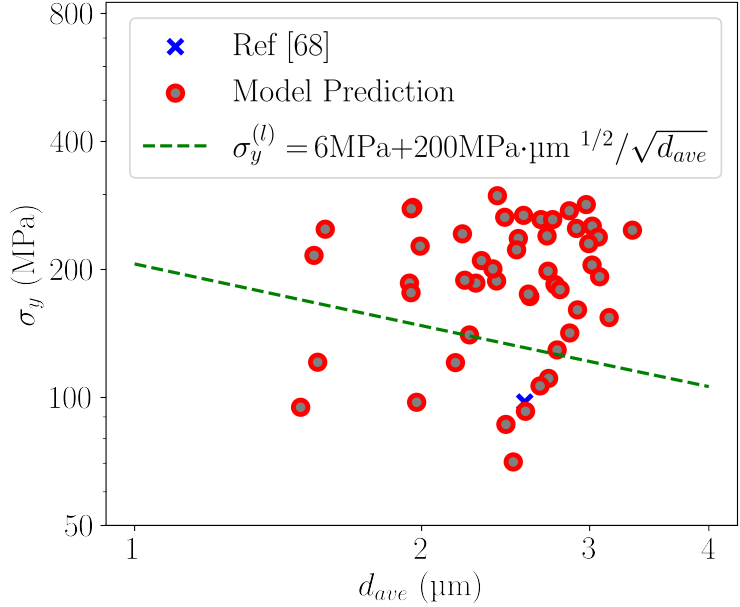

FIG. S2: The predicted yield strength a function of the average grains size, given a bimodal grain size distribution with the peaks at $0.2\,\mu m$ and $12\,\mu m$.

It should also be noted that the two data points of the yield strength from Ref. [66] and one data point of the flow stress ($\varepsilon = 1\%$) from Ref. [74] with average grain sizes below $\leq 0.25\,\mu m$ are abnormally a few hundred MPa smaller than any other published experimental data in literature for a similar grain size and are below the lower bound predicted by our current model. The discrepancy in these data points is most likely associated with unoptimized pulsed electro-deposition processing of the material that led to a non-fully dense material, as well as potential inaccuracies in the mean grain size reported [71], which is beyond the scope of our model. More recent studies on Ni samples processed using optimized pulsed electro-deposition (average grain sizes as low as $120\,nm$) show excellent agreement with our predicted flow stresses [71]. The reference numbers in Sec. S2 follow those in the main context.

5

## S3. COMPARING THE PERFORMANCE OF DIFFERENCE SIZE-EFFECTS LAWS

Fig. S3 shows the performance comparison in the three grain size regimes, $\mathcal{R}_{sub}, \mathcal{R}_{super}$, and $\mathcal{R}_{total}$) for different forms of the flow stress versus average grain size relationships that have been proposed in literature [S17], where $\sigma_0, k_0, \alpha$ and $c$ are all fitting parameters. The linear fitting in the form of $\sigma_f = \sigma_0 - c \cdot d_{ave}$ is considered here as a baseline case. The power law form with three commonly used grain size exponents of $1/3, 1/2$ (i.e., the Hall-Petch relationship), and 1 [S18], are also shown. Several non-power law forms are also shown in Fig. S3. For example, based on Matthew's critical thickness theory [S19], the elastic strain is expressed as a linear function of $\ln d_{ave}/d_{ave}$, and thus the flow stress in the form $\sigma_f = \sigma_0 + k_0 \cdot \ln d_{ave}/d_{ave}$ is also considered. Finally, an exponential form, $\sigma_f = \sigma_0 + k_0 \cdot \exp(-\alpha d_{ave})$, and a logarithmic form $\sigma_f = \sigma_0 - c \cdot \ln d_{ave}$ [S17], are also considered.

The different fitting laws shown in Fig. S3 are compared according to their coefficient of determination, $R^2$, and their mean absolute percentage error, MAPE, which are two common dimensionless curve fitting performance metrics. These are more appropriate here than some other metrics (e.g., root mean square error (RMSE)), due to the orders of magnitude differences in the theoretical flow stresses. Fig. S3 shows that the power law with the grain size exponent around $1/2$ generally outperforms the other forms of size effect laws. Comparing Fig. S3(a) with (c), it is clear that the fitting results in $\mathcal{R}_{super}$ are better than those in $\mathcal{R}_{tot}$, especially at high strains. Nevertheless, the $R^2$ and MAPE of the curve $\sigma_f = \sigma_0 + k_0 \cdot d_{ave}^{-1/2}$ in Fig. S3 for the two ranges are very close, demonstrating that the Hall-Petch relationship is applicable to the entire grain size range. For the small grain size range $\mathcal{R}_{sub}$, the $R^2$ values of all forms of the size effect laws are well below 0.6, while the MAPE values are between 10% and 30%, as shown in Fig. S3(b). Therefore, even though the $R^2$ and MAPE values of the size effect forms $\sigma_f = \sigma_0 - c \cdot d_{ave}$, $\sigma_f = \sigma_0 - c \cdot \ln d_{ave}$ and $\sigma_f = \sigma_0 + k_0 \cdot \exp(-\alpha d_{ave})$ are slightly better than those of the power law forms, there is no arguably best form of the size effect law in $\mathcal{R}_{sub}$, based on those two metrics. This is attributed to the relatively small number of data points associated with the narrow grain size range width (i.e., the span of $\mathcal{R}_{sub}$ is roughly 1 µm). Here, the best fitting curve of $R_{sub}$ is yet to be further investigated.

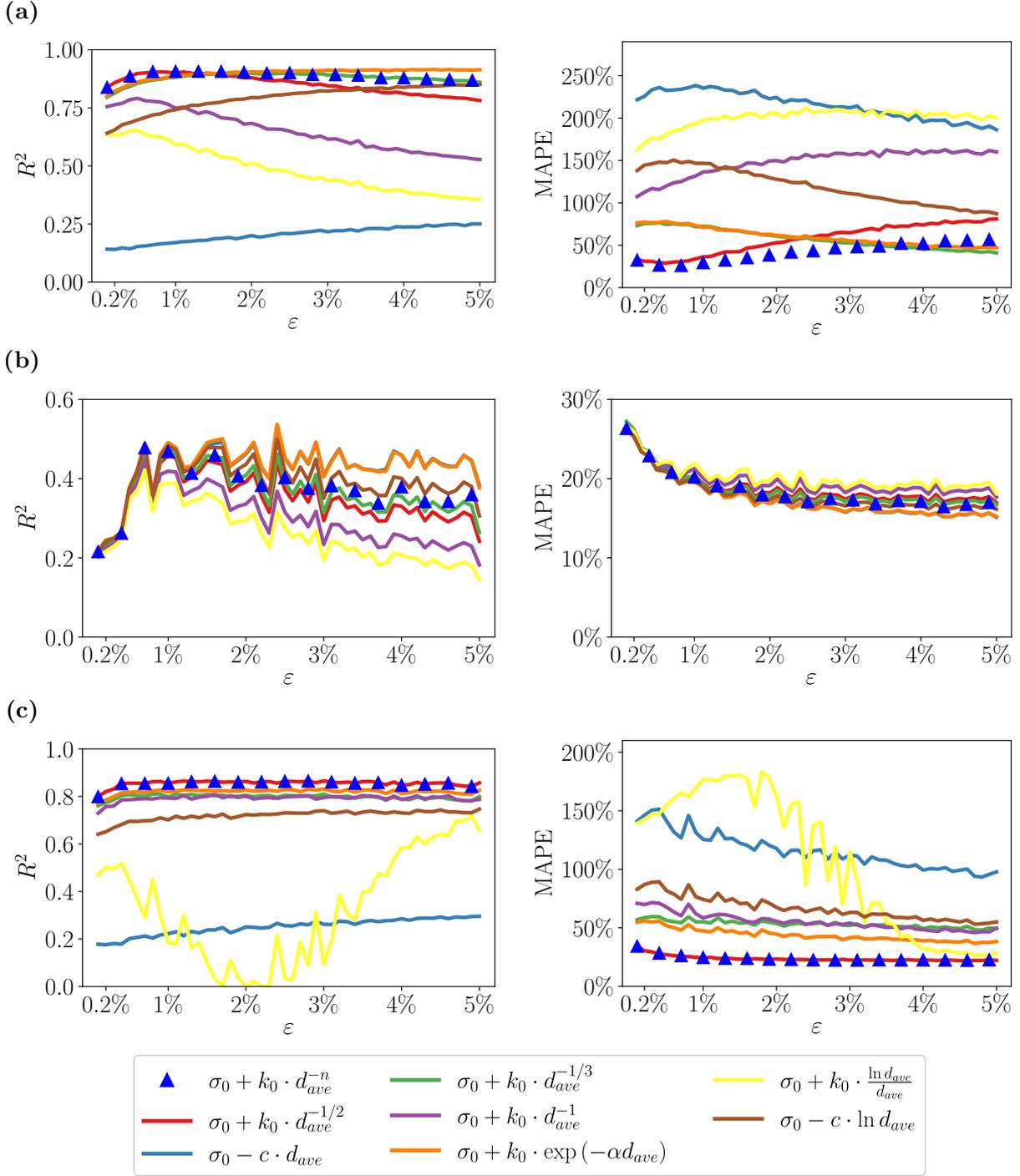

FIG. S3: Performance comparison between different forms of size effect laws with respect to the metrics $R^2$ and MAPE in (a) $\mathcal{R}_{tot}$, (b) $\mathcal{R}_{sub}$ and (c) $\mathcal{R}_{super}$.

## S4. THE MIXTURE DENSITY NETWORK

For any given value of $x$, the probability density function (PDF), $p(t|x)$, of the target data $t$ can be expressed in the form of a Gaussian mixture model :

$$p(t|x) = \sum_{i=1}^{k} \pi_i(x) \mathcal{N}\big(MEAN_i(x), SD_i^2(x)\big), \tag{S1}$$

where $k$ is the number of the components in the model, $\pi_i$ is the weight of the $i$-th PDF with $\sum_{i=1}^{k} = 1$, and $\mathcal{N}(MEAN_i, SD_i^2)$ represents the $i$-th PDF following a normal distribution with the mean of $MEAN_i$ and the standard deviation of $SD_i$. By choosing a mixture model with a sufficient number of components, and a high performance neural network, the MDN can approximate any conditional PDF, $p(t|x)$, with an arbitrary desired error tolerance. The architecture of a mixture density network (MDN) is shown schematically in Fig. S4. The

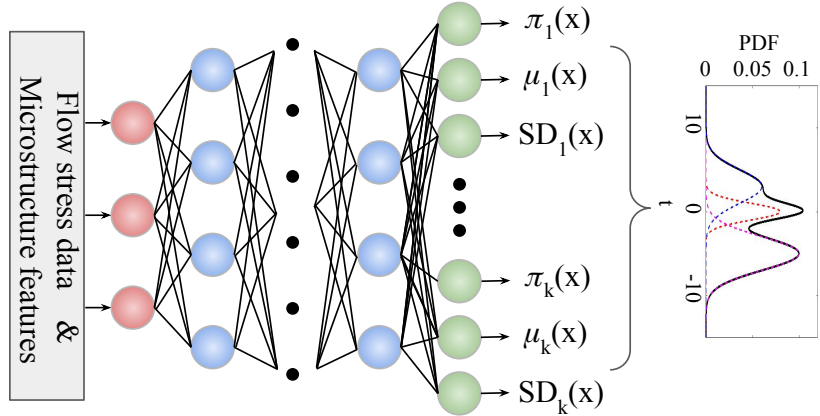

FIG. S4: MDN that emulates a model with three inputs and a one-dimensional output with $k = 3$ mixtures.

loss function is the mean negative log-likelihood, which is defined as

$$\mathcal{L} = -\frac{\sum_x \ln P(X = x)}{\sum_x 1}, \tag{S2}$$

where $X$ is the random variable corresponding to the flow stress of the polycrystalline sample, and $x$ represents all flow stress values occurring in the given dataset.

The input microstructural features into the MDN models were scaled using the min-max scaler, which linearly transforms the input into the range $[0, 1]$, such that the minimum and maximum values of a feature are 0 and 1, respectively. The hyperparameter tuning of

8

the MDN model was performed by training MDN models with different architectures as a process of the grid search among a parameter space, which is listed in Table S3. The number of mixture components is fixed to be 20. The hyperparameter optimization is performed using hyperas where the evaluation metric is the average negative log-likelihood. The model training was performed with a batch size of 64, an epoch of 80, and an early stopping criterion with the patience of 5 epochs.

TABLE S3: The parameter space in MDN training

| Number of hidden layers | Number of neurons in each layer† | Activation function |
|---|---|---|
| 2 | [3,3,3],[5,5,5],[10,10,10]<br>[15,10,5],[20,15,10] | **Sigmoid** or **Relu** or **Elu** |
| 3 | [3,3,3,3], [5,5,5,5],[10,10,10,10]<br>[20,15,10,5],[25,20,15,10] | **Sigmoid** or **Relu** or **Elu** |
| 4 | [3,3,3,3,3], [5,5,5,5,5],[10,10,10,10,10]<br>[30,25,20,15,10,5],[35,30,25,20,15,10] | **Sigmoid** or **Relu** or **Elu** |
| 5 | [3,3,3,3,3], [5,5,5,5,5],[10,10,10,10,10]<br>[30,25,20,15,10,5],[35,30,25,20,15,10] | **Sigmoid** or **Relu** or **Elu** |
| 6 | [3,3,3,3,3,3], [5,5,5,5,5,5],[10,10,10,10,10,10]<br>[35,30,25,20,15,10,5],[40,35,30,25,20,15,10] | **Sigmoid** or **Relu** or **Elu** |

†: Here the layer refers to both the hidden layer and the output layer.

In the analysis of Hall-Petch relationship, as shown in Fig. 3 and Fig. 4 in the main text, the MDN model with the best mean NLL score, has two dense hidden layers, each having 15 and 10 neurons, respectively, while has 5 neurons in the output layer, with the sigmoid activation function.

The feature importance analysis is carried out based on an ensemble of randomly selected datasets and all the MDN model architectures as listed in Table S3. For a given dataset and a given set of features, the corresponding feature importance score, the mean log-likelihood (LL), can be calculated. A larger score means the more important the set of features is in predicting the flow stress. To draw a representative conclusion by reducing the effect of insufficient samples, the feature importance analysis results of different sizes of the subsets

(5%, 10%, 25%, and 100%) of the whole dataset fed into different MDN structures are collected and assembled.

Fig. S5 shows the NLL vs epoch graph for 1000 groups of randomly chosen MDN model used for feature importance analysis, where most of the curves rapidly drop within the first 10 epochs. This shows the efficiency of the MDN models. It should be noted that there are a few curves that only converge to values between 0 and $-0.5$, which is known as a vanishing gradient problem. There are some techniques developed to overcome the vanishing gradient problem, e.g., using ReLu-like activation functions and using batch normalization.

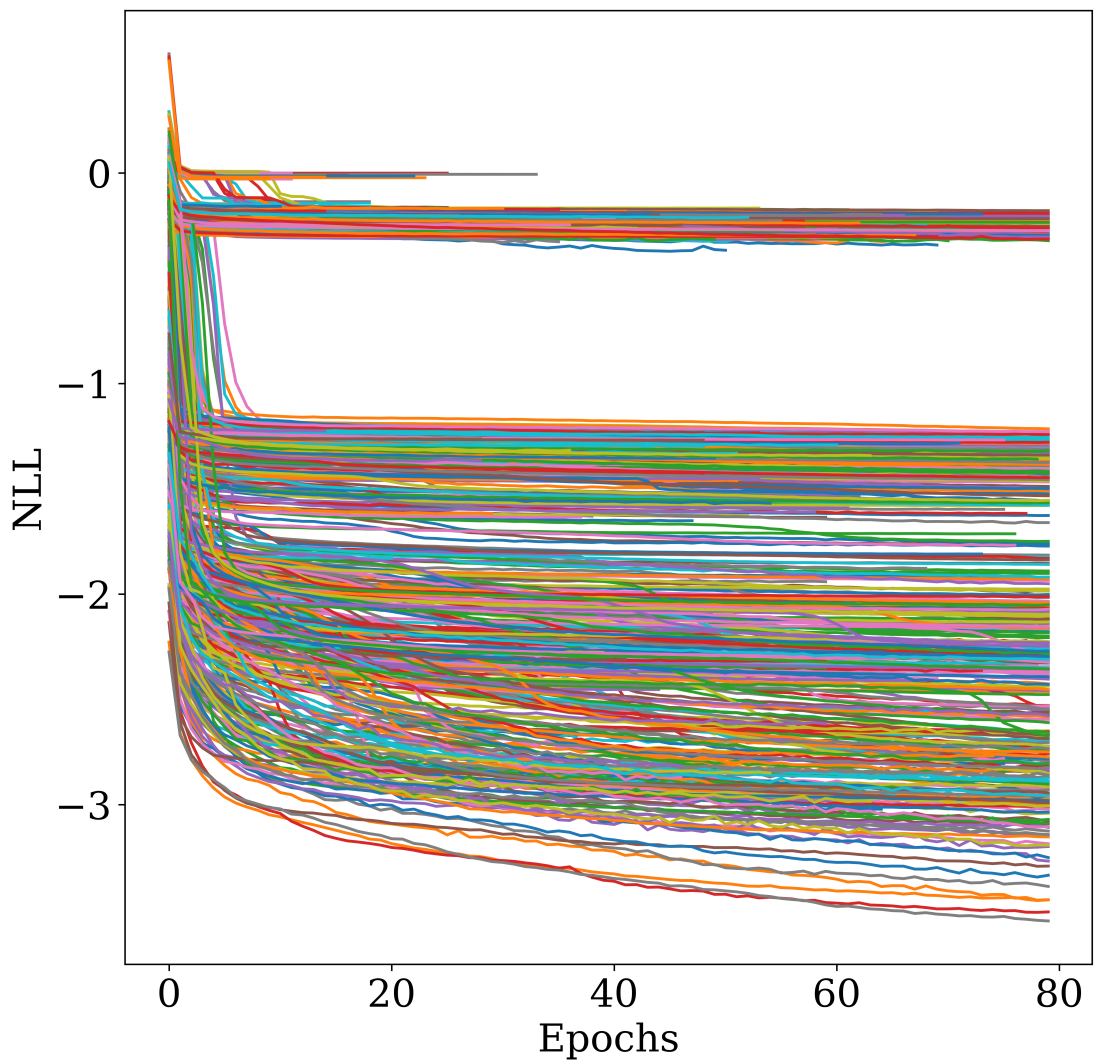

FIG. S5: The NLL vs epoch graph for randomly chosen 1000 trained MDN models.

The theoretical and MDN predicted strength distribution are compared in Fig. S6 using

the metrics of maximum residual error (MRE) and root mean square error (RMSE). MRE and RMSE of $N$ data points are defined as follows:

$$\text{MRE} = \max\{p_i^{MDN} - p_i^{theor} : i = 1, 2, \cdots, N\}. \tag{S3}$$

$$\text{RMSE} = \sqrt{\sum_{i=1}^{N} \frac{\left(p_i^{MDN} - p_i^{theor}\right)^2}{N}}. \tag{S4}$$

To avoid the underestimation of the RMSE due to the vanished relative probability, only the relative probability values larger than a critical value ($1 \times 10^{-4}$ in Fig. S6(b) and $1 \times 10^{-3}$ in Fig. S6(c)) are included in calculating the RMSE. It is shown that the MREs are well below 0.08 and RMSEs are smaller than less than 0.05, which demonstrates the accuracy of the MDN prediction.

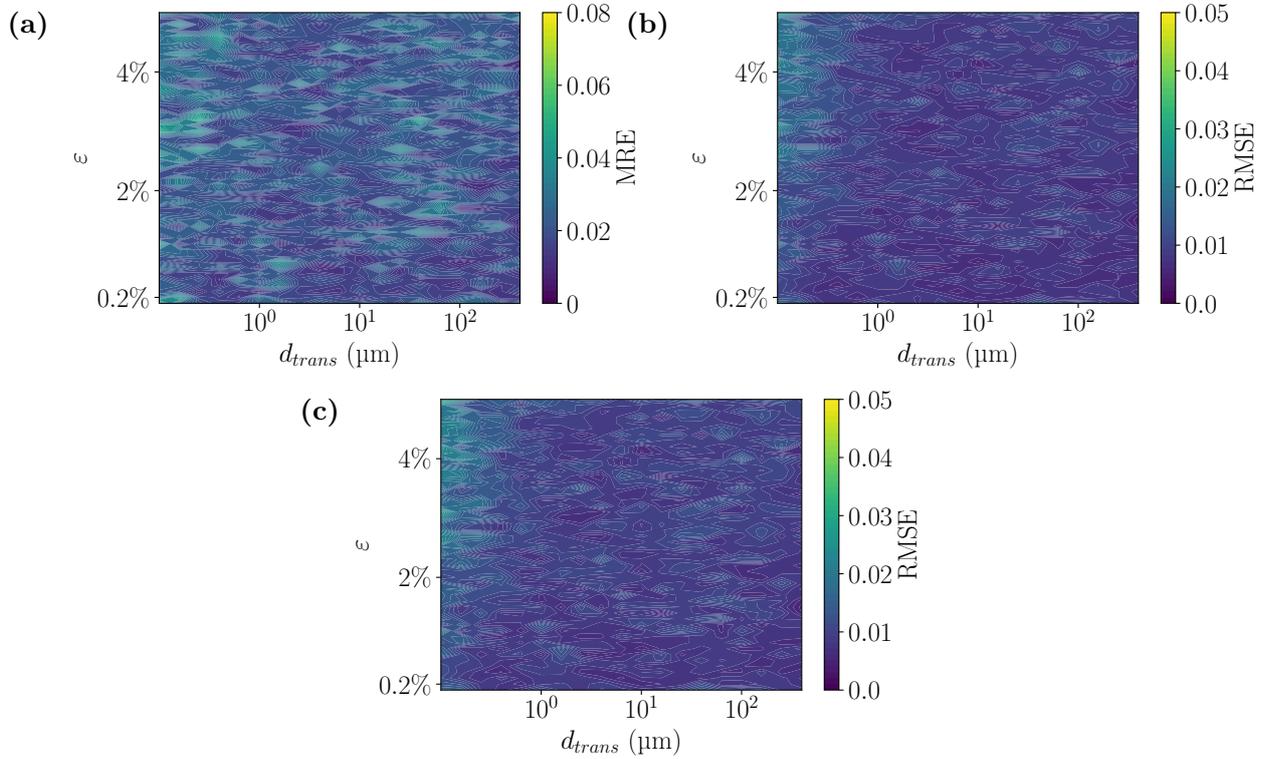

FIG. S6: The contour plot of (a) MRE, (b) RMSE with a critical value $1 \times 10^{-4}$ and (c) RMSE with a critical value $1 \times 10^{-3}$ of the theoretical and the MDN predicted flow stress as a function of average grain size and strain level.

The performance comparison between different forms of curve fitting of quartiles of the theoretical and the MDN predicted flow stress with respect to $R^2$ and MAPE is shown in Figs S7 and S8. Again, the general good alignment of the quartile curves with $1/\sqrt{d_{ave}}$ is

11

evidenced. For quartiles in $\mathcal{R}_{super}$, the $R^2$ is close to 1 and MAPE is close to 0, indicating the power law is a very excellent fit. While for $\mathcal{R}_{sub}$ and large strains, due to the possible lack of data points, the performance of the power law fitting is not as good as those of exponential forms. It should be noted that the quartile analysis performed here is dependent on the strain range discretization. Recently, He et al. developed a rigrous while efficient quartile regression algorithm suitable for large synthetic datasets [S20]. The application of such an algorithm to better understand the quartile behaviors will be presented elsewhere.

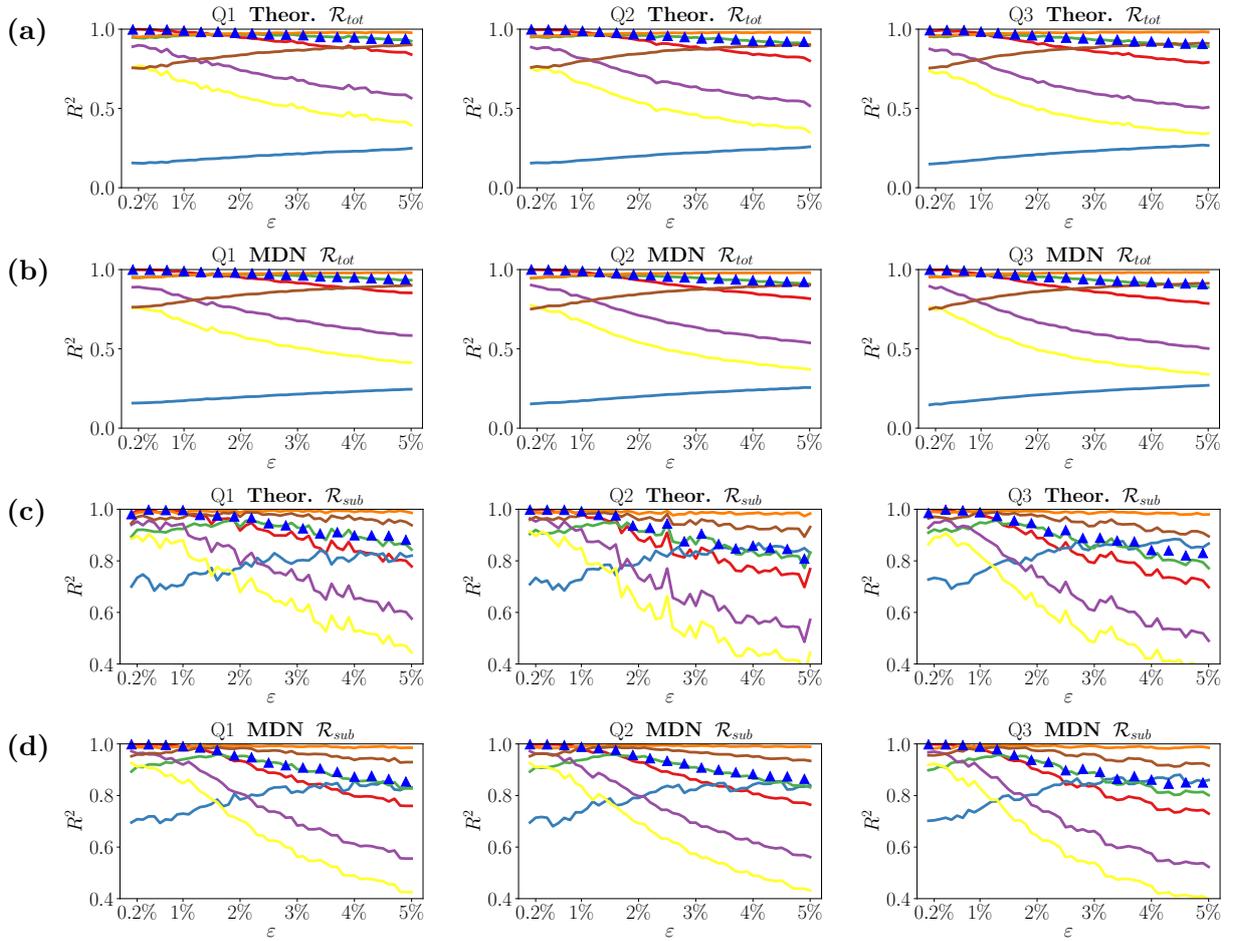

FIG. S7: The performance comparison between different forms of size effect laws with respect to $R^2$. (a) theoretical prediction in $\mathcal{R}_{tot}$, (b) MDN prediction in $\mathcal{R}_{tot}$, (c) theoretical prediction in $\mathcal{R}_{sub}$, (d) MDN prediction in $\mathcal{R}_{sub}$.

The interquartile range (IQR) is defined as the difference between the Q3 and the Q1, that can be used to measure the spread of the flow stresses. The IQR analysis shown in Fig. S9 suggests that in very fine grained and coarse grained materials, the flow stresses are less

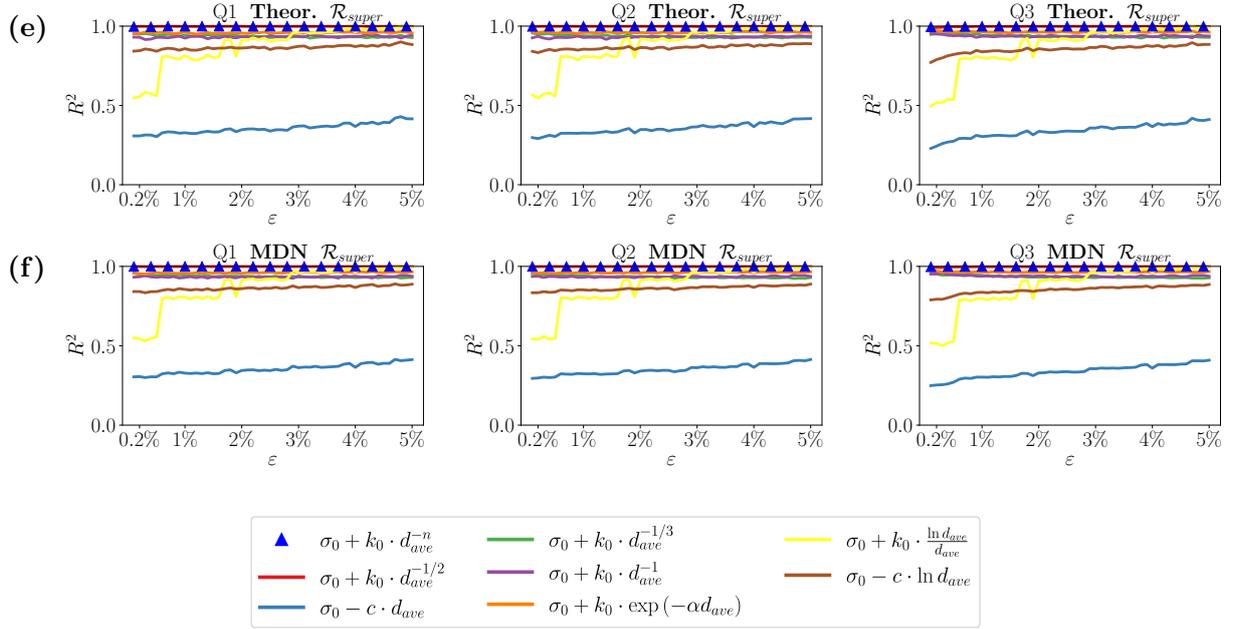

FIG. S7: (Cont'd) The performance comparison between different forms of size effect laws with respect to $R^2$. (e) theoretical prediction in $\mathcal{R}_{super}$, (f) MDN prediction in $\mathcal{R}_{super}$.

dispersed at larger strains, while vice versa in intermediate grain size regimes. This reflects the larger uncertainty associated with more complicated microstructures and deformation mechanisms at large strains. Nonetheless, due to the dominant roles of the theoretical stress in very fine grained materials and the forest hardening term in coarse grained materials, the flow stress scatter associated with large strains is overshadowed.

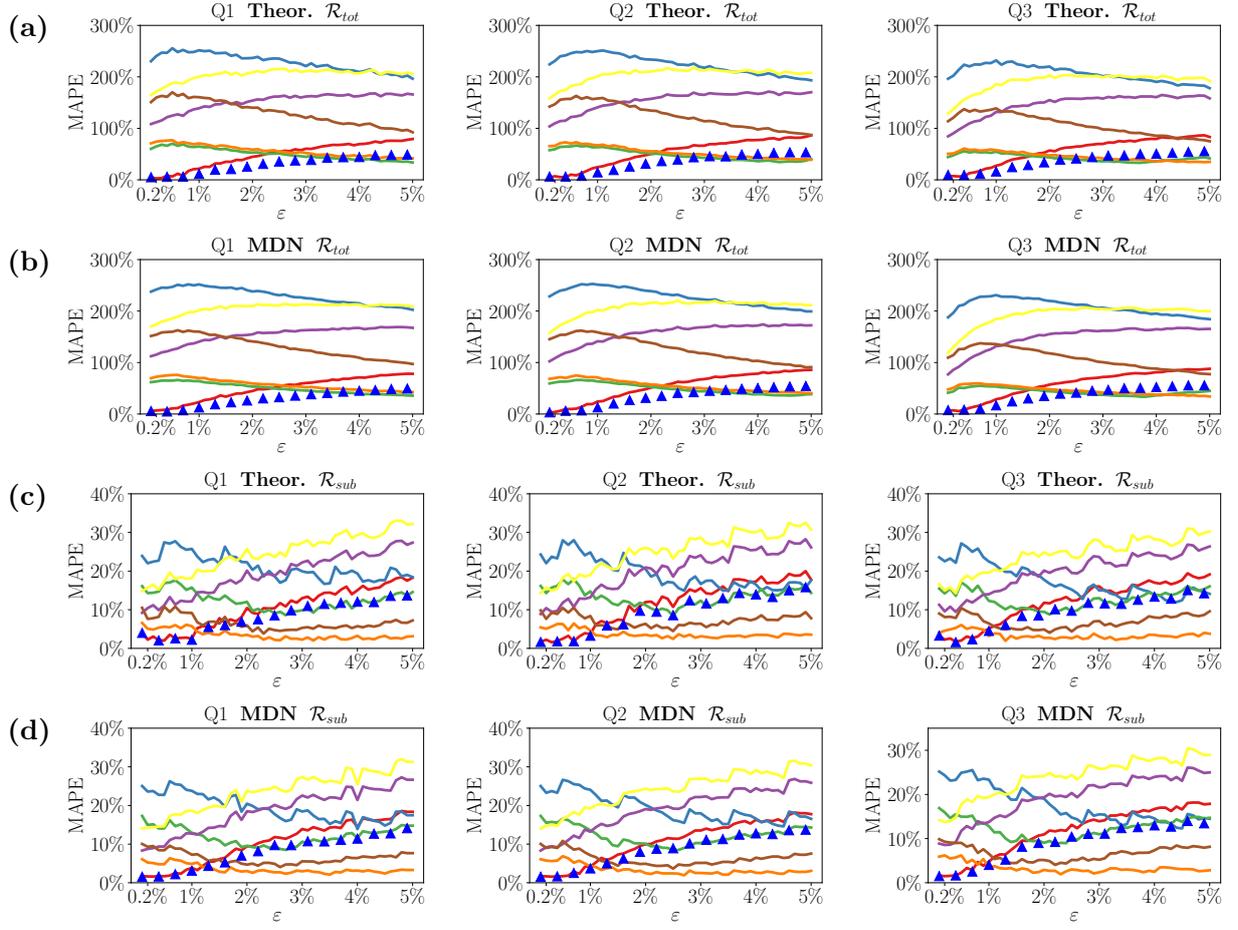

FIG. S8: The performance comparison between different forms of size effect laws with respect to MAPE. (a) theoretical prediction in $\mathcal{R}_{tot}$, (b) MDN prediction in $\mathcal{R}_{tot}$, (c) theoretical prediction in $\mathcal{R}_{sub}$, (d) MDN prediction in $\mathcal{R}_{sub}$.

14

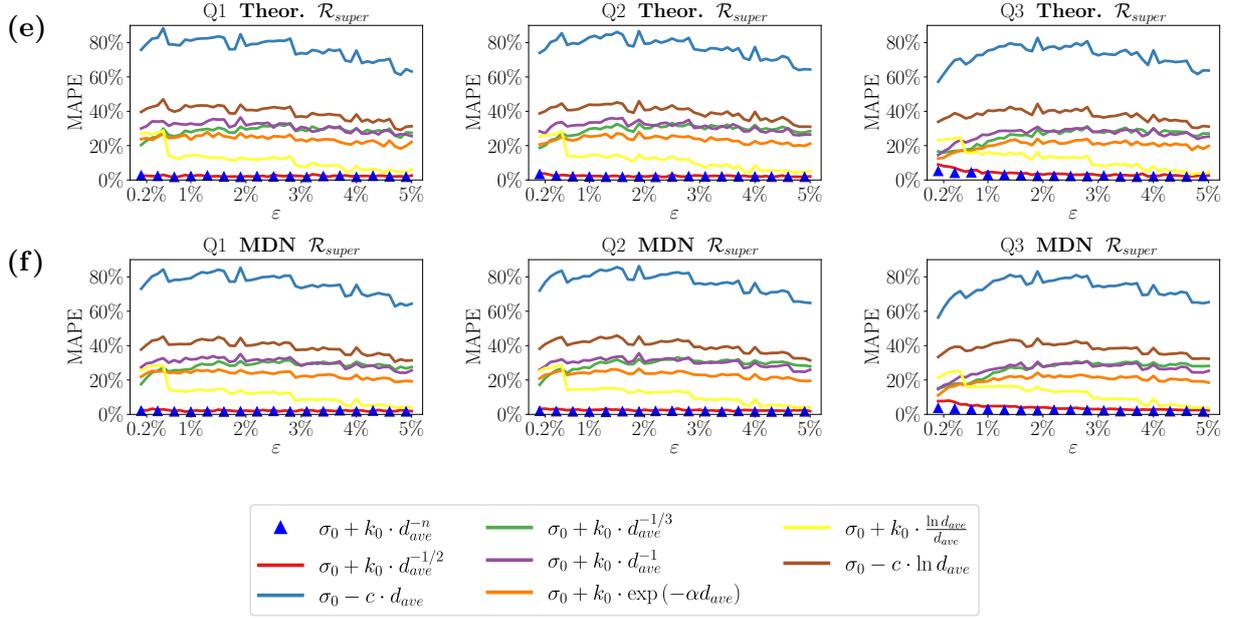

FIG. S8: (Cont'd) The performance comparison between different forms of size effect laws with respect to MAPE. (e) theoretical prediction in $\mathcal{R}_{super}$, (f) MDN prediction in $\mathcal{R}_{super}$.

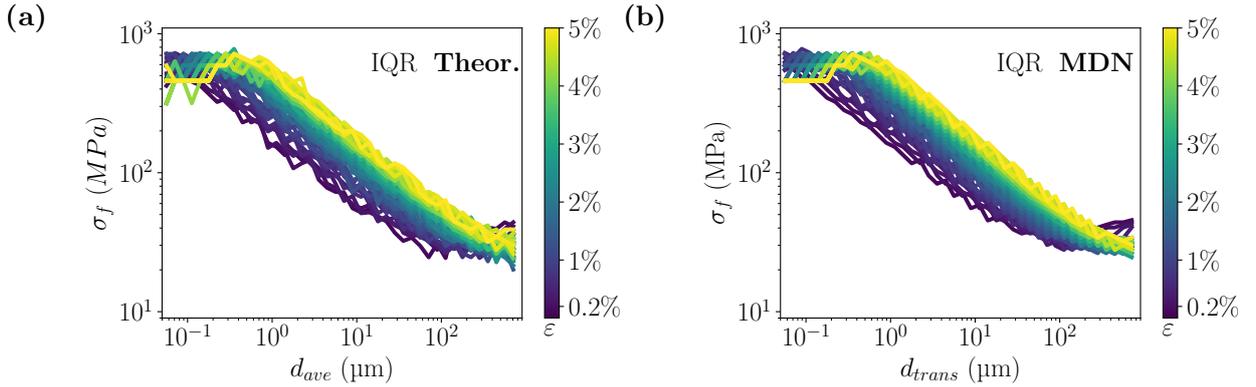

FIG. S9: The IQR of the theoretical and the MDN predicted flow stress for different average grain sizes at different strain levels.



\end{document}